\documentclass{article}

\usepackage{arxiv}

\usepackage[utf8]{inputenc} % allow utf-8 input
\usepackage[T1]{fontenc}    % use 8-bit T1 fonts
\usepackage{hyperref}       % hyperlinks
\usepackage{url}            % simple URL typesetting
\usepackage{booktabs}       % professional-quality tables
\usepackage{amsfonts}       % blackboard math symbols
\usepackage{nicefrac}       % compact symbols for 1/2, etc.
\usepackage{microtype}      % microtypography
\usepackage{lipsum}		% Can be removed after putting your text content
\usepackage{graphicx}
\usepackage[square,sort,comma,numbers]{natbib}
\usepackage{doi}
\usepackage{xcolor}
\usepackage{bm}
\usepackage{amsmath}
\usepackage[toc,page,titletoc]{appendix}
\usepackage[capitalise,nameinlink]{cleveref}
\usepackage{longtable}

\crefname{supp}{Supplement}{Supplements}
\DeclareMathOperator*{\argmax}{arg\,max}  
\DeclareMathOperator*{\argmin}{arg\,min}

\title{Quantifying the evolution of harmony and novelty in western classical music}

\author{ \href{https://orcid.org/0000-0003-2361-1827}{\includegraphics[scale=0.06]{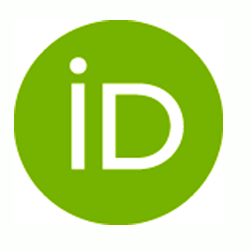}\hspace{1mm}Alfredo Gonz\'alez-Espinoza}\thanks{corresponding author} \\
	Department of Biology\\
	University of Pennsylvania\\
	Philadelphia, PA 19104 \\
	\texttt{jage@sas.upenn.edu} \\
	\And
	Joshua B. Plotkin \\
	Department of Biology\\
	University of Pennsylvania\\
	Philadelphia, PA 19104 \\
	\texttt{jplotkin@sas.upenn.edu} \\
}

\renewcommand{\shorttitle}{Evolution of harmony in classical music}

\hypersetup{
pdftitle={Innovation and evolution of harmony in western classical music},
pdfsubject={},
pdfauthor={Alfredo Gonz\'alez-Espinoza, Joshua B. Plotkin},
pdfkeywords={Complex Systems, Music Information Retrieval, Music Evolution},
}

\begin{document}
\maketitle

\begin{abstract}
Music is a complex socio-cultural construct, which fascinates researchers in diverse fields, as well as the general public. Understanding the historical development of music may help us understand perceptual and cognition, while also yielding insight in the processes of cultural transmission, creativity, and innovation. Here, we present a study of musical features related to harmony, and we document how they evolved over 400 years in western classical music. We developed a variant of the center of effect algorithm to call the most likely for a given set of notes,  to represent a musical piece as a sequence of local keys computed measure by measure. We develop measures to quantify key uncertainty, and diversity and novelty in key transitions. We provide  specific examples to demonstrate the features represented by these concepts, and we argue how they are related to harmonic complexity and can be used to study the evolution of harmony. We confirm several observations and trends previously reported by musicologists and scientists, with some discrepancies during the Classical period. We report a decline in innovation in harmonic transitions in the early classical period followed by a steep increase in the late classical; and  we give an explanation for this finding that is consistent with accounts by music theorists. Finally, we discuss the limitations of this approach for cross-cultural studies and the need for more expressive but still tractable representations of musical scores, as well as a large and reliable musical corpus, for future study.
\end{abstract}

% keywords can be removed

\section{Introduction}

Music represents an important part of our lives, whether for listening or playing, as a hobby or profession.  Even from early societies until today, music is part of everyday life. For this reason, music itself has been a subject of intense study in the fields beyond musicology, including cognitive science, sociology, and even history. A wide swath of academics sett the value of exploring music through an interdisciplinary lens \cite{crosscultural2020,margulis2021}. 

Studies by musicologists on music evolution have provided informative insight and identified trends in music style. While there is recent work in quantitative and statistical analysis of music, most studies have been qualitative. In recent years, the development of technology and digital formats have made the access to music data easier for researchers from different fields, allowing them to address questions ranging from  social interactions and cultural evolution \cite{Savage2020,savage2020b} to creativity or innovation \cite{Jensen2014,Jensen2018,Park2020}. However, even with the access to digital formats, there are still several challenges unique to music, that it apart from studies of other recorded cultural traits such as written language, names, or artistic designs.
%\cite{papers, including canoos/pottery}. 
One of the big challenges in quantifying musical change is to accurately represent music with quantities that are both mathematically and computationally tractable, yet still musicially salient even to non-experts.

Three musical elements are often the focus of analysing musical scores: Melody, Harmony and Rhythm. Although there are many more dimensions to consider, including dynamics and tempo among many others, there is a trade-off between the detail and the dimensionality f the musical representation for purposes of systematic analysis. For example, considering notes from a melody would not include the contextual information from the harmony in the chords. On the other hand, considering all notes in a score in formal analysis will include both melody and harmony but  increase the alphabet size and dimensionality of any representation. While most studies have focused in melodic representation of musical pieces \cite{Liu2010, Dagdug2007c, Gonzalez-Espinoza2020,Su2006,Liu2013,Useche2019,Niklasson2015b,BeltrandelRio2008,Moore2018}, the number of studies considering harmonic properties has increased considerably in recent years \cite{Nardelli2020, Park2020,Telesca2012,Gonzalez-Espinoza2017,Nichols2009,Moss2019, scagliarini2021}. Harmony and tonality are musical concepts that have been exhaustively described by musicians, music theorists \cite{Rameau,Schoenberg,Lewin1987,Lerdahl2001}, psychologists \cite{Temperley2004} and mathematicians \cite{Mazzola2002,Tymoczko2011}. Multiple approaches to quantifying harmony have been implemented, such as defining a measure of consonance \cite{WuBach} for chords \cite{Moss2019,Warrell2020} or individual notes played together (such as codewords) \cite{Park2020,Serra-Peralta2021,Nardelli2020}. We focus our study on  features related to harmony -- to the concept of tonality in particular -- while trying to preserve information relevant in the melody and rhythm. We use a mathematical and computational model for tonality based on  human cognition built around features of a score that have been previously proposed as tonal models, in an effort to preserve as much as musical information as possible based on features that are both scientifically and musically understandable. We provide simple examples of our tonal representations for pieces and composers, 
which illustrate the concepts we define and quantify in this work, in the hope of making the study  accssible o  a broad audience.

\section{Materials and methods}

Reducing dimensionality while preserving information from a  musical score can be achieved by defining a feature of a higher = order, other than the series of singular notes in the score itself. One possibility is to consider chords instead of individual notes, as most  music theorists do in harmony analysis \cite{Moss2019,Neuwirth2018}). This approach requires defining a mapping from sets of notes to chords -- e.g. the notes C-E-G maps to the C chord. But developing algorithms to re-label set of notes as chords in a systematic fashion for a large corpus is difficult without a good definition or metric of harmony, and it doesn not account account for note duration or rhythm.

One alternative approach is to define a {\em local key} given a set of notes in a contiguous region of the score, athough this is still an open problem lacking clear defitions. Different algorithms have been developed aiming to approach the problem of local key identification \cite{Temperley1999WhatsKF,Krumhansl1991,Dawson2017}. We implement a variation of the {\em center of effect} algorithm previously introduced by E. Chew \cite{Chew2002}, which has been used for music information retrieval problems \cite{Chew2005-2,Chew2005, Herremans2019} and has been shown to outperform other algorithms, compared to gold-standard hand annotations by musicologists \cite{Chew2014}. 

\subsection{Key representation in the spiral array}
\begin{figure}[h]
	\centering
    \includegraphics[width=0.7\textwidth]{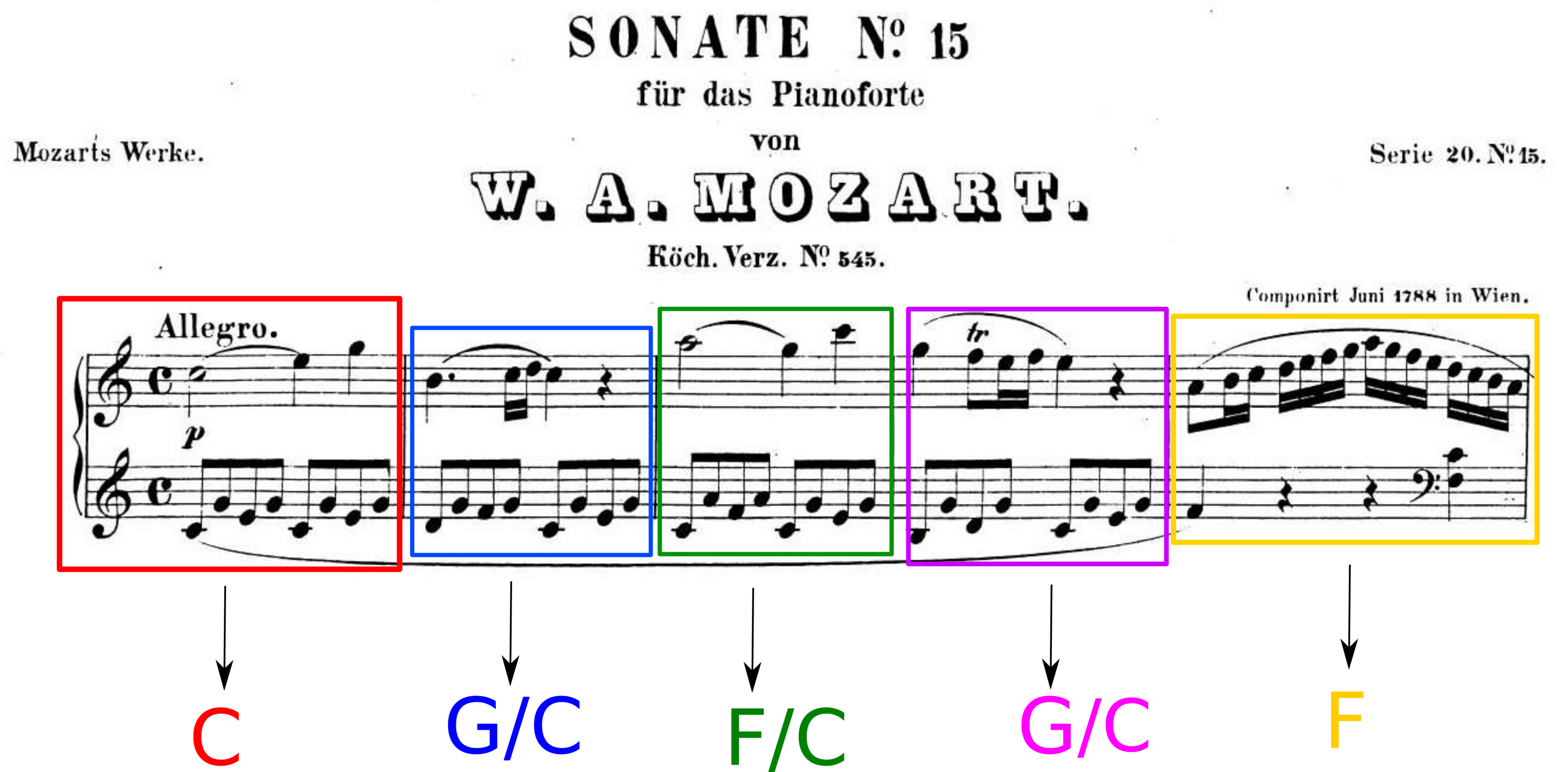}
    \caption{Example of determining the ``local key" of each measure. The figure shows the first five measures from   piano sonata K. 545 in C by W.~A.~Mozart, along with the most likely key (or two most likely keys) determined by the CoE algorithm..}
	\label{mozart}
\end{figure}

We will represent a musical score as an ordered sequence of elements: $\xi = \{ e_1,e_2,...,e_n\}$, where each element corresponds to the local key of each measure (or fixed number of measures) in the piece (see figure \ref{mozart}). We use a geometrical representation where notes, chords and keys are represented as points $(x,z,y) \in \mathbb{R}^3$ in a helix-type configuration known as the spiral array \cite{Chew2014}.

Using the spiral representation and center of effect as model for tonality allows us to not only analyze a large amount of musical score but also to relate their mathematical features with concepts from information theory. Derived from the tonnetz network on Riemann's theory \cite{LHiggins1962-1, LHiggins1962-2, Tonnetz1997}, the spiral representation preserves the hierarchical structure of tonality since key representations are generated from combining chords, and chord representations from combining notes while representing all of these structures in the same  space ($\mathbb{R}^3$).

Notes in the spiral array are defined as 
\begin{equation}
    \vec{P}(k) = \begin{bmatrix} x_k \\ y_k \\ z_k \end{bmatrix} = \begin{bmatrix} r sin\frac{k\pi}{2} \\ r cos\frac{k\pi}{2} \\ k h\end{bmatrix} ,
\end{equation}
where $r$ and $h$ are parameters (see SI) and $k$ is a number representing a specific note. The starting note $k_0$ is chosen arbitrary, for simplicity we define $k_0$ as the C note.
%This number $k$ is assigned arbitrary, by simplicity we define C $= 0$. 
Notes $k$ and $k+n$ are separated by $n$ fifths, preserving the harmonic relationships between notes, chords and keys (e.g. with $k_0=$ C, $k+1=$ G, $k+2=$ D and so on). Major and minor chords ($\vec{C}_M$ and $\vec{C}_m$) are constructed by linear combinations of notes and major and minor keys ($\vec{T}_M$ and $\vec{T}_m$) from linear combinations of chords (see SI).

The {\em center of effect} is based in Von Neumann's center of gravity (mass) where a set of notes $P = \{\vec{p}_1, \vec{p}_2,...,\vec{p}_N\}$ has an effective center of mass in the form of a linear combination of its elements
\begin{equation}
    \vec{C}_e = \sum_{i=1}^N \omega_i \vec{p_i} \label{coe1}.
\end{equation}
Here the coefficients $\omega_i$ are normalized ($\sum \omega_i = 1$) and they represent the {\em importance} of each note. These coefficients can be defined in multiple ways (see supplementary \ref{coe_def}) but we choose to use the normalized duration of each note -- so that our representation captures some aspects of rhythmic structure. The Center of Effect (CoE) key finding algorithm we develop uses the vector $\vec{c}_e$ for the set of notes, and defines the most likely key as:
\begin{equation}
    \argmin_{T \in \bm{T}} || \vec{c}_e - \vec{T} ||,
\end{equation}
which corresponds to the key $T$ for which the euclidean distance to the measure is minimum. Here $\bm{T}$ is the set of possible major and minor keys: $\bm{T} = \{ \bm{T}_M(k) \forall k\} \cup \{\bm{T}_m(k) \forall k \}$. The CoE method also assigns a likelihood to the most likely local key, based on the euclidean distance above, as well as a likelihood to all alternative keys based on their distances (Eq.~\ref{eq:keylikelihood}). 

The CoE algorithm has proven to be effective and to have better performance than other methods when defining a key with small amount of information \cite{Chew2014}, and has shown useful not only to identify a key but to other applications like pitch spelling \cite{Chew2005-2} and passage segmentation even with post-tonal music from composers like Messiaen\cite{Chew2005}. By construction, the spiral representation is en-harmonic in-equivalent, meaning that it distinguishes between sharp and flat notes that  equal temperament would consider to be the same (e.g. C\# and Db). This in most cases can be an advantage of the model.
%, however in our case for convenience we require for it to be enharmonic equivalent. 

We evaluated the accuracy our implementation of the CoE (Center of Effect), including the various weights we introduce in its implementation, by comparing its output to a gold standard of local keys annotated by hand, measure by measure, for Beethoven's string quartets  \cite{Neuwirth2018}. This annotated data is one of the most detailed set of such information in the literature, it contains information about the key and chord in functional harmony for each measure in all Beethoven's string quartets. Our CoE implementation matches the hand annotation in for more than 68.43\% of the measures in the corpus, and cases of mis-matches are often musically plausible alternatives (see SI).

\subsection{Key diversity and uncertainty}
We use two quantities to describe harmonic features in a musical piece: key diversity and key uncertainty. Key diversity represents how diverse  the distribution of most-likely keys are, across measures in a score. For each piece $\xi$ we compute the probability for each of its elements as their normalized frequency (counts) in the sequence:
\begin{equation}
    p(e_i) = \frac{f(e_i)}{\sum_{e_j \in \xi} f(e_j)},
\end{equation}
where the sum is over the elements in the sequence $\xi$. The key diversity of a piece is related to how many different keys the piece contains, over measures, and how localized or spread  out their probability distribution.

Key uncertainty is a quantity that gives us information about how {\em ambiguous} is the tonal reference. The more ambiguous the less certain we are about calling the  key in each measure. The center of effect algorithm returns a list of possible keys $\{T_1, T_2, T_3, ..., T_m\}$ with their associated distances $\{d(T_1), d(T_2),..., d(T_m)\}$ to the center that represents the notes ($\vec{C}_e$).  We define $p(T_i)$ as the probability for a set of notes to be in the key $T_i$ as a function of the distance $d(T_i)$: 
\begin{equation}
\label{eq:keylikelihood}
    p(T_i) = \frac{e^{- \lambda d(T_i)}}{\sum_j e^{- \lambda d(T_j)} },
\end{equation}
where the parameter $\lambda$ is defined by setting $P(T_i) > 0.98$ for the case when $d(T_i) \approx 0$. The closest the key $T_i$ is to $\vec{C}_e$ the most likely key given the set of notes in $\vec{C}_e$ represented in that key. If the $\vec{C}_e$ is similarly close to more than one key, their probabilities are similar to the most likely key, and key uncertainty would be high.

We use Shannon's definition of entropy, also known as the expected information content for a random variable $X$, defined as $H(X) = - \sum_{x \in X} p(x) log(p(x))$ to compute both key diversity (across measures) and key uncertainty (within each measure) with their respective probability distributions $p(e)$ and $p(T)$. In the case of the key uncertainty, we perform one computation for each measure of the piece and we take the average uncertainty of the piece over all measures in the piece: \\
\begin{equation}
    \text{Key Uncertainty} = - \frac{1}{M} \sum_{m=1}^M \sum_{T \in \bm{T}} p(T) log(p(T)),
\end{equation}
where $M$ is the total number of measures in the piece.

\subsection{Innovation in key transitions}
Although notes in a musical score are known to have long range dependencies \cite{Voss1978, Gonzalez-Espinoza2017}, several studies have used memory-one Markov chains  to quantify musical scores \cite{Park2020,Moss2019,Verbeurgt2004}, where the state $s_{t+1}$ in a musical sequence depends only on the previous state $s_{t}$ or a set of previous states $\{ s_{t},s_{t-1}...\}$).Markov chains of a higher order representations, such as {\em local keys}, viewed as harmony transitions, are perhaps more useful than Markov Chains on individual notes that lack of contextual information \cite{Moore2018}. 

A piece $\xi = \{ e_1,e_2,...,e_n\}$ can be seen at the outcome  a Markov Chain where the states are the elements (local keys) the transitions between elements are given by a stochastic matrix $\bm{P}_{\xi}$ with the initial state $e_1 \in \xi$. The value of the entry $P_{\xi ij}$ correspond to the probability of the transition between the states $e_i \rightarrow e_j$. Given an empirical score, represented as a series of local keys, the Markov transition matrix is computed via maximum likelihood estimation:
\begin{equation}
    P_{\xi ij} = P_\xi(e_i \rightarrow e_j) = P_\xi(e_j | e_i) = \frac{f_{\xi}(e_i,e_j)}{ \sum_{x \in V_\xi}f_{\xi}(e_i, x)},
    \label{prob_tran}
\end{equation}
where $f_{\xi}(e_i,e_j)$ is the frequency (counts) for the bi-gram $(e_i, e_j)$ in the piece $\xi$ and the term $\sum_{x \in V_\xi}f_{\xi}(e_i, x)$ is equivalent to the frequency of the element $e_i$ since it is over all the elements $x \in V_\xi$, where $V_\xi$ is the set of unique elements (vocabulary) in $\xi$.

One of our goals is to evaluate how {\em innovative} a given piece $\xi$ is, when compared with a set of previous historical works. This comparison is made by assuming that a model representing all previous works $\Omega$ can be used to reproduce $\xi$. We quantify how how good the model $\bm{P}_\Omega$ is at generating the piece reproduce $\bm{P}_\xi$. We can directly relate this quantity to the Kullback-Leibler divergence, which is a measure for the amount of extra information needed to code a distribution $P$ from a given distribution $Q$.
For stochastic matrices the $KL$-divergence rate is defined by (see SI):
\begin{equation}
     D_{KL_R}(\bm{P}_\xi||\bm{P}_{\Omega}) = \sum_{e_i \in V_\Xi} \sum_{e_j \in V_\Xi} \mu_{i} P_\xi(e_j|e_i)\cdot log\left(\frac{P_\xi(e_j|e_i)}{P_{\Omega}(e_j|e_i)}\right),
    \label{kld_eq}
\end{equation}
where $\mu_i$ corresponds to the asymptotic distribution of elements in $\xi$. Here, $D_{KL_R}(\bm{P}_\xi||\bm{P}_{\Omega})$ is the amount of information per step the model $\bm{P}_\Omega$ needs to reproduce $\bm{P}_\xi$. This quantity is different from information content used in previous works to quantify novelty \cite{Park2020} since it considers the asymptotic distributions of the elements $\mu_i$ (the more repetitive the less novel) and does not depend on the length of the sequence (see fig. \ref{app:kld_size}). 

To quantify innovation for a given piece from year $y_i$, we construct a transition matrix by adding all key transitions from pieces in the previous years 
\begin{equation}
    \bm{P}_{\Omega_i} = \hat{\bm{F}}_{\Omega_i} \quad \text{with} \quad \bm{F}_{\Omega_i} = \sum_{y_j < y_i} \bm{F}_{\xi_j}
\end{equation}
where $\hat{\bm{F}}_{\Omega_i}$ is the normalized frequency matrix or stochastic matrix for all the pieces in the previous years. We then compute the Kullback-Leiber divergence as in Eq.~\ref{kld_eq} and refer to this value as how novel (the novelty value) of a piece, given all preceding scores.

\subsection{Corpus}
We use a set of MIDI files from the Kunstderfugue \cite{kunstderfugue} website. This data set have been used in previous studies \cite{Gonzalez-Espinoza2017,Gonzalez-Espinoza2020,Park2020,Serra-Peralta2021}, it consists of a compilation of $~18,000$ MIDI files from more than 79 western composers, from the years 1200-1950. We retained only those pieces to which we could assign a year of composition (see SI), reducing the dataset to 5,374 MIDI files. We processed the pieces using a Julia script \cite{Agonzal2023} to extract the information needed to compute the center of effect -- such as numbers of measure, pitch, and duration of all notes (see supplementary material for details).  

\section{Results}
To test the performance of our implementation of the center of effect model, we compared the results to hand-annotations of local key  performed by musicologists, obtaining an accuracy of $>68.43 \%$ across all measures in Beethoven's string quartets \cite{Neuwirth2018}. Some cases of mis-matches against this gold standard are necessarily incorrect example of calling the local key (see supplementary \ref{app:supp4}). Overall, the ability to match hand annotations this often seems to validate our methodology, especially consider that professional musicians and musicologists may disagree about the key of a measure, even within the well structured Beethoven quartets. It is somewhat useful, in addition, that the CoE algorithm provides several alternatives, with associated likelihoods, for each measure -- because the entropy of this distribution provides a measure of key uncertainty, which is itself an interesting feature in music.

\subsection{Key diversity and uncertainty}
\begin{figure}[h]
	\centering
    \includegraphics[width=.7\textwidth]{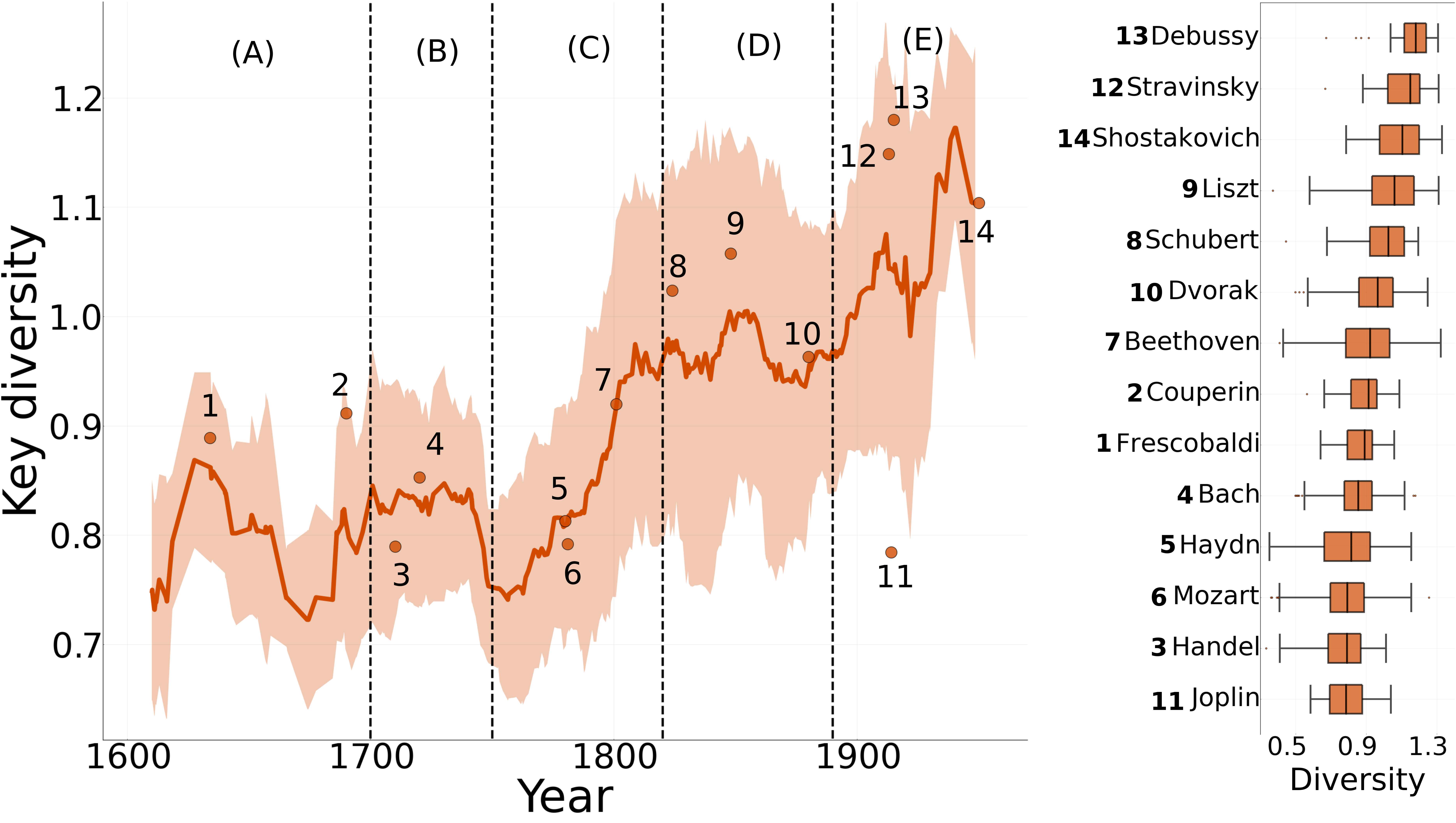}
    \caption{{\bf Key diversity.} Time series for key diversity across 5,374 musical scores. The solid line represents the median of the distribution of values within a time bin, while the first and third quartiles are represented by the shaded area. All distributions are computed for values that lie within a sliding window of 20 years. Historical periods are denoted by dashed lines, where the letters stand for: (A) Early/Mid Baroque, (B) Late Baroque, (C) Classical, (D) Romantic and (E) Modern.}
	\label{key_div}
\end{figure}

A time series show key diversity (within each piece) is  shown in figure \ref{key_div}. We observe a trend towards increasing diversity of keys within a piece, over time, which is consistent with  results from previous studies that show how many elements in music tend to be more diverse over time \cite{Serra-Peralta2021}. In our case, key diversity does not follow a constant increasing trend. Rather, the the change in diversity is different in specific periods of time associated roughly with different periods of classical music defined by musicologists (Early/Mid Baroque, Late Baroque, Classical,  Romantic,  and Modern).  
The increase in key diversity is most evident during the Classical (1750-1820) and Modern (1890-1960) periods. This result in the Classical period agrees with the historical analysis in the style of music \cite{Rosen1998}. Indeed, the early Classical period saw the development of new instruments that increased the size of the orchestras, allowing composers  o explore more tonal modulation and musical forms. A similar trend is seen in the Modern period, which is characterized by more experimentation of modulation and more complex conceptualizations of tonality or even the rejection of tonality altogether \cite{Berg1930}. 

\begin{figure}[h]
	\centering
    \includegraphics[width=.7\textwidth]{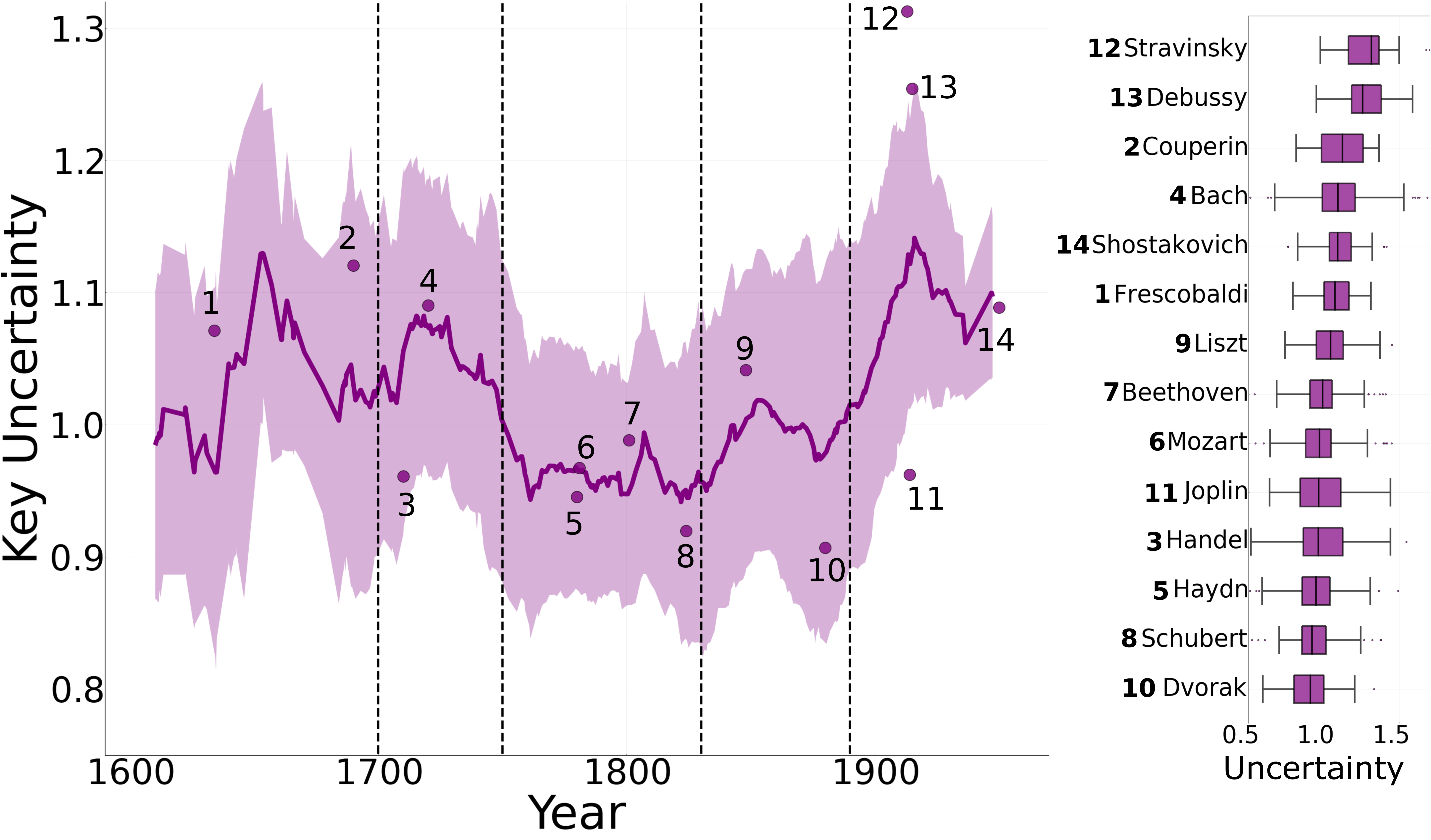}
    \caption{{\bf Key uncertainty.} Time series for the key uncertainty value, the solid line represents the median of the distribution of values while the first and third quartiles are represented by the shaded area. Regions between dashed lines are the historical periods presented in figure \ref{key_div}}
	\label{key_unc}
\end{figure}

Figure \ref{key_unc} shows the timeseries of key uncertainty over the 5,374 scores in our dataset. The overall trend shows an increase the uncertainty associated with assigning local keys to measures by the CoE algorithm, over the course of 400 years. This increasing trend in key uncertainty may reflect changing concepts of tonality. Key uncertainty can be interpreted as how ambiguous a tonal center (key) is, and so ambiguity may be related harmonic complexity or immediate tonal tension \cite{Herremans2016}. A secular changing in key uncertainty are largely concordant with  historical features of the Baroque and Classical periods \cite{Kamien2008}. For instance, in the Baroque period, music was not only polyphonic but also contrapunctual, promoting higher density of different notes and more propensity to dissonances. In the early Classical period the texture of the sound is clearer than in the Baroque period \cite{Kamien2008}, giving more emphasis to order and hierarchy resulting in a homophonic texture with a clear melody above a chordal accompaniment that results in a less complex representation of tonality. 

\begin{figure}[h]
	\centering
    \includegraphics[width=0.7\textwidth]{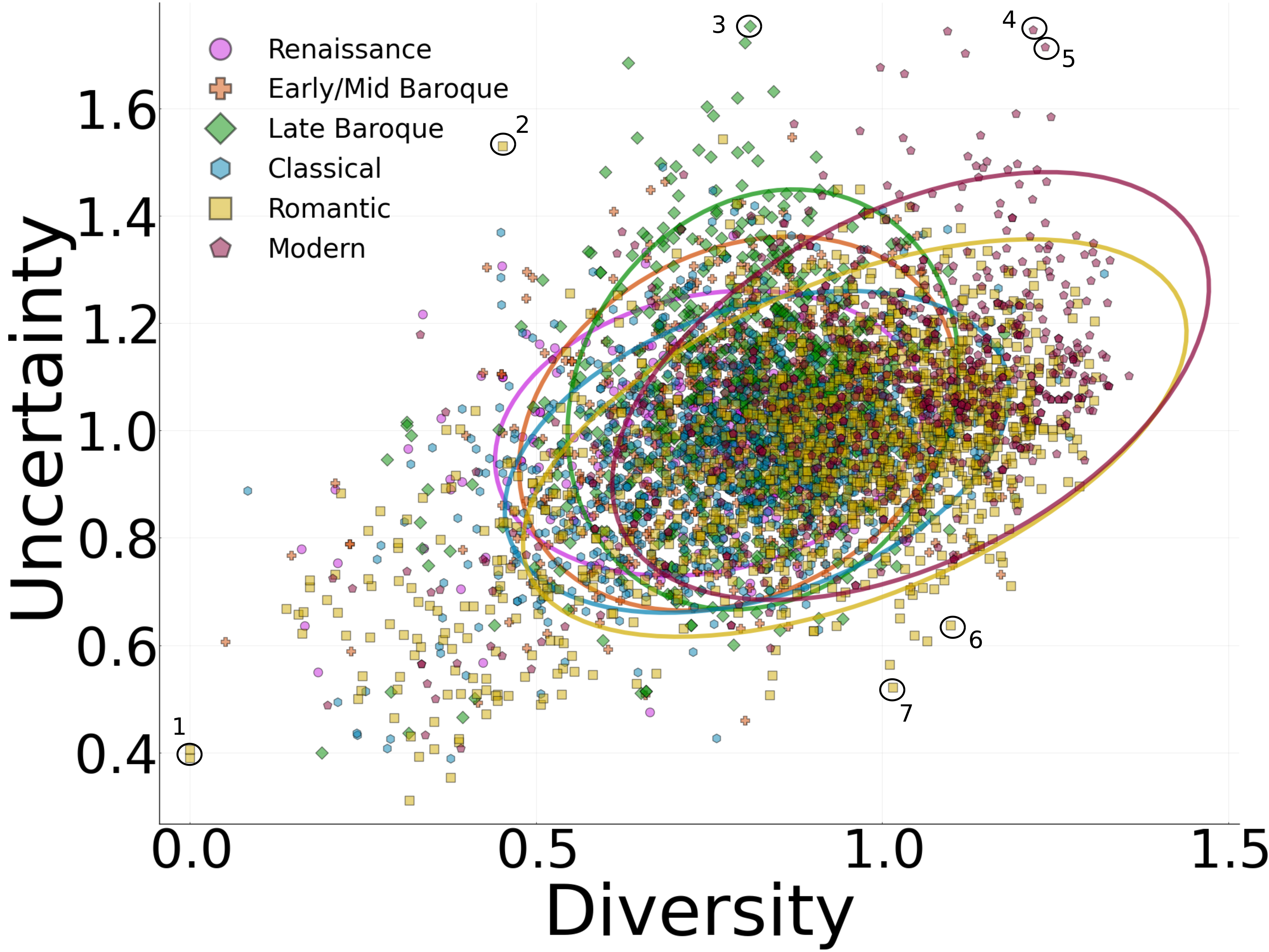}
    \caption{{\bf Key Diversity vs Key Uncertainty.} Diversity and uncertainty values for individual pieces classified by their historical period. The ellipses are constructed by computing the covariance error with 95\% confidence. }
	\label{div_unc}
\end{figure}

While key diversity and key uncertainty represent distinct aspects of a musical piece, the 400-year trend of overall increasing in value holds for both, with the modern period showing the highest uncertainty and diversity. The relationship between diversity and uncertainty is shown in figure \ref{div_unc}, with a Pearson correlation value of $\rho = 0.38$. In the same figure some of the extreme scores are selected and highlighted in circles, and these particular pieces are listed in table \ref{comp_table}.  

Simply listening the pieces highlighted in Figure \ref{div_unc} -- which are selected for having extreme values of key diversity and/or key uncertainty -- helps provide an intuitive understanding of what these two measures quantify. And so we include audio files for each pieces in table \ref{comp_table}, representing low uncertainty - low diversity, low uncertainty - high diversity, high uncertainty - low diversity and high uncertainty - high diversity extremes across the corpus. For example, the pieces that are most uncertain and most diverse in tonality are typically from modern composers (Scriabin and Stravinsky), while  the other extreme (least uncertain and least diverse)  corresponds to a simple religious hymn (The Great Physician) that was written to be easy to sing and to remember. 
\begin{table}[h]
\centering
\begin{tabular}{|c|c|c|c|}
\hline
Piece & Year & Name                                & Composer              \\ \hline
1     & 1869 & The Great Physician                 & William Hunter / Hymn \\ \hline
2     & 1868 & 30 Progressive Etudes - Etude \#1   & Joachim Raff          \\ \hline
3     & 1720 & Two part inventions - Invention \#5 & J.S. Bach             \\ \hline
4     & 1911 & 3 Etudes Op 65. - Etude \#1         & Scriabin              \\ \hline
5     & 1910 & Capture of the Firebird (excerpt)   & Igor Stravinsky       \\ \hline
6     & 1880 & Waltzes 54 - \#6 in F Major         & Antonin Dvorak        \\ \hline
7     & 1828 & Sonata 959 Mov 3                    & Franz Schubert        \\ \hline

\end{tabular}
\caption{{\bf Compositions List.} List of the seven pieces highlighted in figure \ref{div_unc}, the number in the first column corresponds to the label in figure \ref{div_unc}.} \label{comp_table}
\end{table}

\subsection{Innovation in key transitions}
We compute a novelty value for each piece using the Kullback-Leibler divergence rate from equation \ref{kld_eq}, described in more detail in the materials and methods section. The novelty value is computed with two alternative key representations: the original key or the transposed key. The original representation key refers to the actual name of the key in the measure (C,E,D etc) while the transposed corresponds the key relative a given reference--in our case the reference is the tonic or the main key of the piece. We use these two different key representations to control for the fact that tonic keys of pieces underwent substantial change over the course of the dataset. The transposed novelty measure attempts to control for this secular variation in tonic keys, when computing novelty.

For the  representation of novelty after transposition, we use roman numerals, as in harmony analysis (see \ref{fun_har_sup} for details), maping every key sequence to a sequence of roman numerals in which I denotes the global tonic key of the piece. This mapping allows us not only to study sequences of roman numerals (functional harmony), but to transpose all the pieces and analyze them within the same reference, avoiding transitions that have the same harmonic relation to be counted as distinct.  

\begin{figure}[h]
	\centering
    \includegraphics[width=0.7\textwidth]{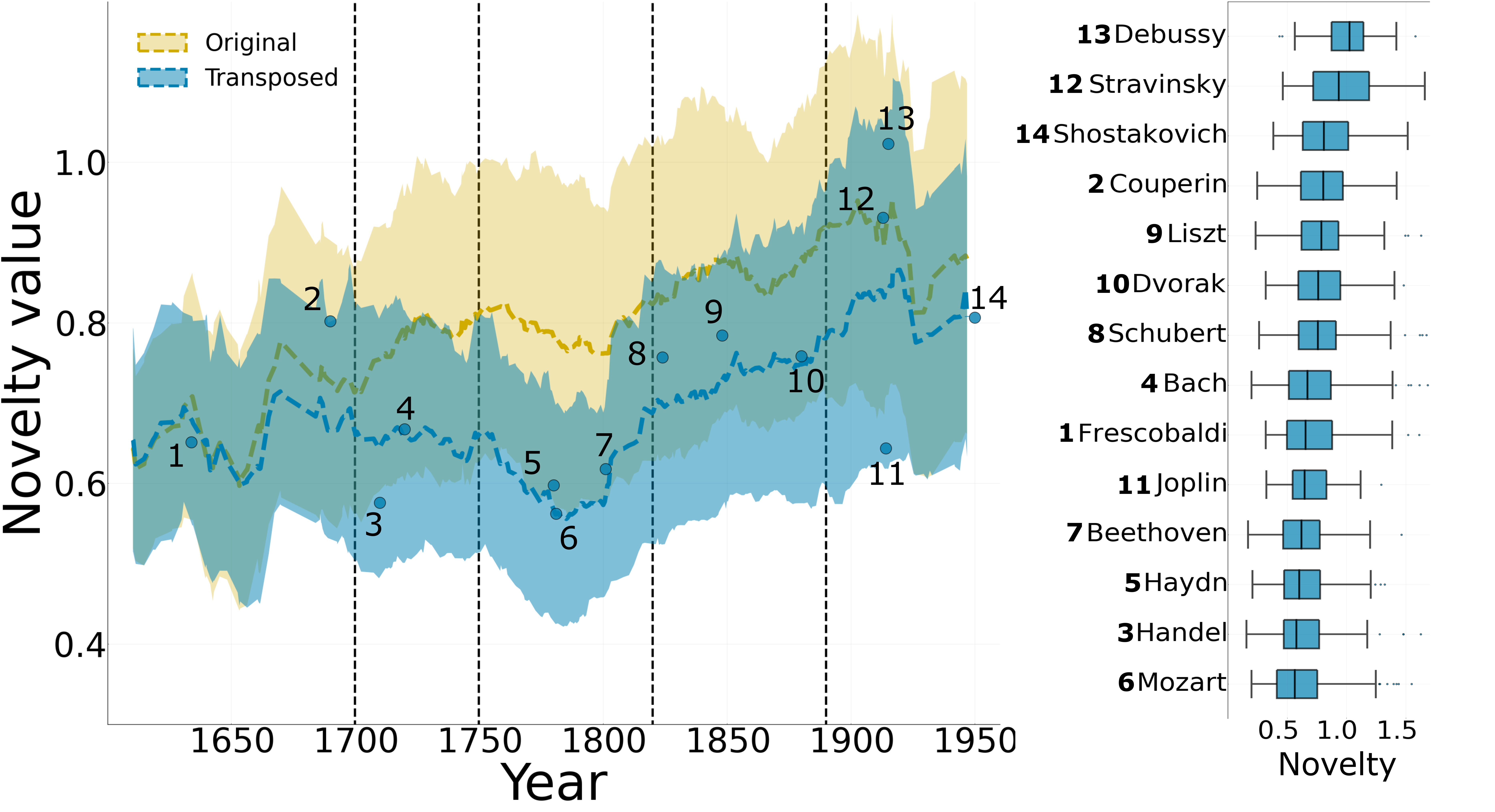}
    \caption{{\bf Innovation values.} Novelty or innovation values for original key representation (yellow) and transposed or functional harmony representation (blue). Values are computed in the same fashion as in figure \ref{key_div} with an overlapping window of 20 years.}
	\label{inov_all}
\end{figure}
Results for  innovations values in both original and transposed representations are shown in figure \ref{inov_all}. Lower innovation values overall for transposed pieces are expected, e.g. in the case for the transition E $\rightarrow$ A in a piece with tonic key E, the transition becomes I $\rightarrow$ IV, and is the same as in any other major piece where the tonic transitions to the subdominant, making the piece in E less novel than it would be scored in the original key representation. 

Figure \ref{inov_all} shows a notable decline in novelty or innovation of harmonic transitions in the Classical period. Although perhaps surprising, this  feature has been previously reported, based on a smaller set of pieces and with a representation that considers as different chords with the same harmonic function  \cite{Park2020}. Our results, by contrast, show that the decline in innovation for harmonic transitions during the classical period is even more evident after transposing all pieces to the same key -- that is, focusing on harmonic transitions per se, after accounting for any possible changes in the tonic key of the piece. 

\begin{figure}[h]
	\centering
    \includegraphics[width=0.9\textwidth]{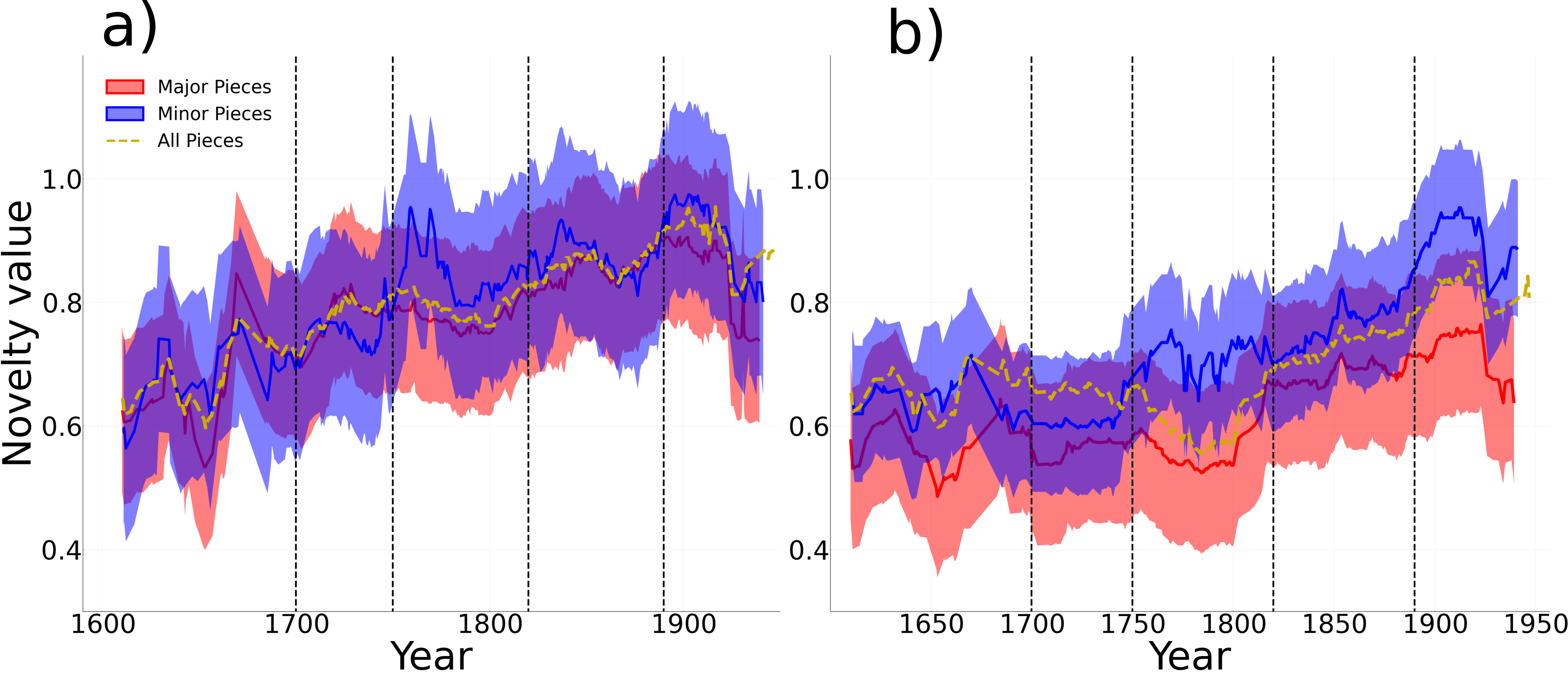}
    \caption{{\bf a) Original keys.} Novelty values for scores represented in their original keys (without transposition), separating minor and major pieces.{\bf b) Novelty scores after transposition to a common key.} }
	\label{inov_funhar}
\end{figure}

Surprisingly, one of the least innovative composers according to this measure is W.~A.~Mozart. But it is well known that Mozart's style, like Haydn's, is an archetype of the classical style, where clarity, balance and transparency are the characteristics of his work \cite{Cliff2006, Heartz2003}. It was until his late period of active years when he exploited chromatic harmony, one example is his popular String Quartet in C major also known as the ``Dissonance" quartet.  Other observations can be made from the results in figure \ref{inov_all}, such as the change of novelty between the classical and romantic period, where Beethoven (number 7, indicated in the figure) played an important role in the movement that  consolidated with composers including Schubert and Liszt (numbers 8 and 9 respectively). The increasing trend in harmonic transition innovation after the second half of the Classical period is a tipping point that led to subsequent continued increase in innovation in the Romantic and Modern periods, which are known for their development in musical form and harmony representation. 
%
%\textcolor{red}{(working on it...)}The difference in the novelty results between the original and transposed pieces in the late Classical period, can be explained without transposing to the same key of reference, can be understood by common key transitions that occur in a uncommon global tonal key. For example, one of the most common transitions; C $\rightarrow$ C,  when the piece is in C major it has a very small contribution to novelty value, but  a larger value when the piece tonal key of the piece is, say, E major.
%%%
We also computed separately novelty values for pieces in major and minor keys (see fig. \ref{inov_funhar}). We found that although the trends in both major and minor tonalities are similar, they differ considerably in the functional harmony (transposed) representation with a clear separation in the classical period. This could indicate that the exploration of modulation occurred more often in minor keys. This result may reflect the fact that minor tonalities have three different scale patterns compared with major tonalities, with only one; and these three patterns prove a richest space of harmonic modulation in a minor key compared to a major key.

%Trying to guess the elements of a piece with a low novelty value is not a difficult task, we know that given our definition of novelty the more repetitive a piece is and the more common its elements are, the less novel or innovative it will be.  

\begin{table}[h]
\centering
\begin{tabular}{|c|c|c|c|c|}
\hline
Year & Name                        & Composer     & Novelty value & First 10 elements       \\ \hline
1570 & O Sacrum Convivium          & F. Guerrero  & 0.2137        & I-IV-I-ii-I-IV-I-I-V-IV \\ \hline
1696 & String Sonata II Mov 4      & D. Buxtehude & 0.1969        & I-I-IV-I-I-I-I-I-IV-I   \\ \hline
1749 & Royal Fireworks Suite Mov 4 & G.F. Handel  & 0.1550        & I-I-I-I-V-V-V-V-I-I     \\ \hline
1798 & 7 Lander in D               & Beethoven    & 0.1670        & I-I-I-V-I-I-I-V-I-I     \\ \hline
1867 & Les Mois Op. 74 Mov 4       & Alkan        & 0.1888        & I-I-I-I-I-I-i-I-I-I     \\ \hline
\end{tabular}
\caption{{\bf Compositions with lowest novelty values.} List of compositions with the lowest novelty values on each period of time, and their first 10 elements. Full sequences are included in supplementary material (\ref{full_seqs})}\label{low_inov}
\end{table}
A list of the five compositions with the lowest novelty value for each period of time is presented in table \ref{low_inov}. The elements of the key sequences in these pieces are very repetitive the most well known in music theory, such as the tonic (I), the dominant (V) and the subdominant (IV). 

%For the case of ``most innovative" compositions the task of trying to guess what elements these would have is more challenging since these elements would need to be rare in the previous pieces and the probabilities in equation \ref{kld_eq} become less trivial to guess. 

Examples for the pieces with highest novelty values for each period are listed in table \ref{hi_inov}. Inspecting the local keys of the first 10 measures of these pieces we find many examples that do not appear among the common local keys in  table \ref{low_inov}, with the exception of the piece by Schubert that starts with common keys before changing to more rare keys (see \ref{full_seqs}).
\begin{table}[h]
\centering
\begin{tabular}{|c|c|c|c|c|}
\hline
Year & Name                          & Composer    & Novelty value & First 10 elements                                            \\ \hline
1570 & Madrigals Book 4 \#19         & C. Gesualdo & 1.72          & I-IV-I-vi-V-IV-I-IV-ii-I                            \\ \hline
1696 & Mass for the parishes Mov 14  & F. Couperin & 1.8597        & I-i-i-i-v-IV-I-i-I-IV                               \\ \hline
1718 & WTC I Prelude 14 in F\# minor & J.S. Bach   & 2.1341        & i-i-\#VI/bVII-i-\#VI/bVII-\#II/bIII-\#II/bIII-v-v-v \\ \hline
1817 & Sonata Op. 147 Mov 1          & F. Schubert & 1.6387        & I-I-I-V-I-I-I-I-I-V                                 \\ \hline
1861 & Esquisses Op. 63 No. 7        & Alkan       & 1.58          & I-I-I-I-vi-\#i/bii-v-\#i/bii-III                    \\ \hline
\end{tabular}
\caption{{\bf Compositions with highest novelty values.} List of compositions with the highest novelty values on each period of time and their first 10 elements. Full sequences are included in supplementary material (\ref{full_seqs})}\label{hi_inov}
\end{table}

%\begin{figure}[h]
%	\centering
%    \includegraphics[width=0.8\textwidth]{Figures/all_profiles.png}
%   \caption{Composer corpus $P_{C}$ Sliding window $P_{\Omega_w}$  }
%	\label{inv_prof}
%\end{figure}

\section{Discussion}

%\subsection{The center of effect: advantages and limitations}
The center of effect model has been shown to identify a meaningful notion of local key, in both tonal and even  non-tonal music \cite{Chew2005}. Our implementation does a decent job of matching a gold standard corpus of manual annotations by musicians. The extended application of CoE for local key calling may be useful not only to musicians who want to explore artistic development, but also to the scientific community that is interested in the quantitative description and development of  variation in tonality and tonal tension, within a piece or across different pieces or time periods.  

One disadvantages of  this approach is the need to select a window size where the center of effect is computed. We have chosen a window size of one measure, but this is a somewhat arbitrary decision and, indeed, measure length shows variation over time and within pieces. The local key may change in a meaninful sense even within; or variation across two or three measures may also be meaningful. We can see this issue in the example shown in figure \ref{mozart}, where some measures may be assigned to two plausible different keys (e.g. in measure \#2 there are two possible keys: C and D); and we choose the most likely one according to its distance to the center of effect. Addressing this complixity requires a different approach with automatic segmentation, where the location for the change in key is determined by a new parameter that can be expressed as a function of the distance between different centers of effect calculated within different time windows (\cite{Chew2002}). Such an  implementation would ideally reduce the dispersion for the uncertainty and diversity values from figures \ref{key_div} and \ref{key_unc}, but this remains a topic for future technical development.

%\subsection{Key uncertainty and diversity}
The structure and evolution of rhythmic and melodic  properties have been explored previously in several studies on large corpora \cite{BeltrandelRio2008,Levitin2012c,Telesca2012,Gonzalez-Espinoza2020, Gonzalez-Espinoza2017,Dagdug2007c}, mostly using time series analysis methods to describe concepts like scaling, predictability and reversibility within a given musical piece.  

By contrast to most of the time-series methods used in these studies, we have focused our analysis by coarse-graining a musical score, but in a way that hopefully retains specifically musical information. That is, we represent a piece as a series of transitions between local keys, from measure to measure.%and that could be of easier interpretation for the general audience, we define the concept of {\em local key} that can be understood as the immediate tonal reference within a time frame. 
Using our definition of local key we can attribute information-theoretic concepts to musical pieces, such as  diversity and average uncertainty, based on an intrinsically musical representation of the piece. In particular, key uncertainty is a concept we have used to quantify the degree of uncertainty whe n a set of notes is assigned a key; and this uncertainty can also be related to how unpredictable is, a given set of notes, to know what note will be added next. Similar types of unpredictability has been quantified in \cite{Gonzalez-Espinoza2017}, where it was shown an overall increasing trend for more unpredictability of notes over time, with Shostakovich being one of the most unpredictable composers in short time frames. Those findings agree with  our results for secular trends in key uncertainty, which again finds  Shostakovich amongst amoung composers with highest key uncertainty. These results are also consistent with the trend for the development of tonality, which it was not formally defined until 1722 in Jean Phillip Rameau's work {\em Treatise on harmony} \cite{Rameau}. The decline of uncertainty during the classical period  reflections a period of time when this convention became adopted by many composers, changing the texture of the polyphonic and contrapunctual forms of  the Baroque period to more clear and homophonic forms defined by a melody and subordinate chordal accompaniments. The subsequent increase of uncertainty in the Romantic and Modern periods reflects the evolution of the concept of tonality in the late Romantic and early modern periods with composers like  Bruckner, Tchaikovsky, Mahler, Scriabin, Wagner and Strauss, with the last two being known for ``furthered the musical language of Opera taking tonality itself to breaking point".  \cite{Wildridge2022}.  Schoenberg described this period of tonality as "fluctuating" or "suspended", implying that it was not decided or it was ambiguous \cite{Schoenberg}. Finally, in the Modern period harmonic progressions become more unpredictable making tonality even more ambiguous, as Meyer describes ``the increased use of the ambiguous chords, the less probable harmonic progressions, and the more unusual melodic and rhythmic inflections ... the felt probabilities of the style system had become obscure; at worst, they were approaching a uniformity which provided few guides for either composition or listening" \cite{Meyer1967}. 

Key diversity within a piece also has a close relationship with tonality, as it is closely related to harmonic function (functional harmony). In this case  the trend from the Classical, Romantic and Modern periods largely coincide with trends  in  uncertainty. 
%The descriptions by Schoenberg and Meyer can be used in the same way here, since the more diverse a piece is the more difficult to define a {\em tonal center}, then the concept of tonal center fades and harmonic transitions become more unpredictable. 
But one important differences between uncertainty and diversity is that the increase in diversity is more evident in the late Classical period, indicating  increasing modulations within a piece. This result agrees with the historical fact that in the late Classical period composers started to explore different harmonic transitions and innovations, as  the industrial revolution contributed to the expansion and diversification of orchestras giving more material to composers to explore different styles and sounds\cite{Brian2002,Downs1992,Rosen1972}. In the scientific literature there are few studies are directly related to key uncertainty and diversity whereas more  studies explore the distribution of notes, chords or other tokens in analogy to language. There is a substantial body of work  establishing empirical laws like Zipf's and Heap's laws in such musical features. One such study considers the concept of vocabulary richness, considering an element (or token) as the set of different notes played during a beat; and the authors find an increasing linear trend in vocabulary richness over time \cite{Serra-Peralta2021}.

%\subsection{Harmonic innovation}
Our study shows the Classical period as a tipping point for novelty in harmonic transitions. While this does not means that there was no innovation prior to this period, it is a quantitative account of the evolution of harmony discussed in qualitative terms by many musical historians   \cite{Meyer1967,Brian2002,Downs1992,Rosen1972}. The evident increase in key diversity or modulation is a component that plays an important role in our computation of novelty. However, key diversity is not the only factor that contributes to novelty, as recurrent transitions in a piece reduce its novelty alue under our metric. This means that even if a piece has novel transitions, if the piece is also highly repetitive it will not be score as novel as if it had fewer repetitions. This effect can be seen in the value for the entropy rate for each piece that appears to not have a particular trend over time (see supplementary figure \ref{ent_rate}), meaning that some music has a functional repetitive component. Indeed, Schoenberg describes music as a perfect balance between repetition and surprise\cite{Schoenberg}. 

Our is  not the first study trying to quantify novelty in musical scores. In \cite{Park2020} the authors address a similar question, although  they represent a piece a sequence of chords without making any attempt to determine the local key of a measure.  \cite{Park2020}  was also constrained to  a considerably smaller corps ($<1000$ pieces) and fewer composers (~20). (And, notably, sometimes all of the pieces by a given composer were assigned the same year, in \cite{Park2020}. Although ours results share some similarities with \cite{Park2020},  it also has its notable differences. For instance, Clementi is one of the least novel composers according to \cite{Park2020}, while Mozart is the least novel in our study. B
%eethoven shows higher in novelty than Schubert while in our case is the opposite, and one can find other significantly different results. 
Both approaches show  is a decline of novelty during the Classical period, although is not clear if the  explanation is the same, because the lack of transposition and reduction of octaves in \cite{Park2020} produce a fundamentally different novelty measure, that is influenced by variation in the tonic of pieces. We believe that  transposition are important to consider features of harmonic relevance: a chord played on a higher octave is not more novel than on a lower one, in our analysis (unlike   \cite{Park2020}); and the  the same piece played in a different key is not considered novel according to our meaure, unlike  \cite{Park2020}. 
%be considered as a different piece, with our methodology we avoid these type of issues that could lead to different results (as compared in figure \ref{inov_all}). 

Although we have been able to quantify novelty in harmonic transitions, and identify historical trends, much remains for a systematic understand of the underlying process for innovation in music. The mechanisms that allows composers or musicians to create new transitions could involve microscopic details, for example, to define a new local key (center of effect) the linear combination of its elements can be modified by adding another note or simply modifying the coefficients and this change could end up in a different key. This hypothesis shares some similarities with ideas about the emergence of innovations by  correlated novelties, or the ``adjacent possible"  \cite{Tria2014,Iacopini2017}. In those models, 
a new state (local key in our case) is discovered by combining previous ideas that are {\em semantically} close enough. Future work to identify actual processes of  innovation would be valuable, where the elements are not only local keys but also combinations of notes, rhythmic patterns, melodic and harmonic motifs and dynamics. 

%\section{Conclusion}
%We combined a computational approach to tonality previously used particularly as a tool for musicians and musicologist with information theory concepts to characterize and quantify harmonic aspects in western classical music from 1500 to 1920. Using a high hierarchy representation of harmony such as the local key, we were able to identify historical trends in the representation of tonality and its diversity. 

\section*{Acknowledgements}
We thank the Plotkin lab for engaging discussion and feedback,  Vladimir Viro for providing fruitful feedback.
%and XXX and for for assistance in data handling. 

\section*{Author's contribution}
A.G-E. and J.B.P. designed the study and developed the methodology. A.G.-E. processed and curated the data. A.G.-E. implemented the programming code, performed the calculations and compiled the results. A.G.-E. and J.B.P. analyzed and interpreted the results. A.G.-E. and J.B.P. wrote and 
revised the manuscript.

\section*{Competing interests}
The authors have declared no competing interests.
\clearpage
\newpage

\bibliographystyle{unsrt}
%\bibliography{references}
%\bibliography{apssamp}
 
\clearpage
\newpage

\renewcommand{\thepage}{S\arabic{page}} 
\renewcommand{\thesection}{S\arabic{section}}  
\renewcommand{\thetable}{S\arabic{table}}  
\renewcommand{\thefigure}{S\arabic{figure}}
\renewcommand{\theequation}{S\arabic{equation}}

\setcounter{section}{0}
\setcounter{page}{1}
\setcounter{equation}{0}
\setcounter{table}{0}
\setcounter{figure}{0}
\begin{center}
    \Large{\textbf{Supplementary Material}}
\end{center}

\section{Spiral representation: definitions and parameters}
We use the Spiral representation introduced by E. Chew \cite{Chew2014} on which each pitch, chord and key is represented by a point in $\mathbb{R}^3$. This representation is given by fifths: a pitch $k$ and a pitch $k+n$ are separated by $n$ fifths. 
Pitches are defined as:
\begin{equation}
    \vec{P}(k) = \begin{bmatrix} x_k \\ y_k \\ z_k \end{bmatrix} = \begin{bmatrix} r sin\frac{k\pi}{2} \\ r cos\frac{k\pi}{2} \\ k h\end{bmatrix} ,
    \label{pitchrep}
\end{equation}
where $r$ and $h$ are fixed: $r=1$ and $h=(2/15)^{1/2}$. The major and minor chords are constructed as linear combinations of pitches:
\begin{equation}
    \vec{C}_M(k) = w_1 \vec{P}(k) + w_2 \vec{P}(k+1) + w_3 \vec{P}(k+4),  
\end{equation}
and
\begin{equation}
    \vec{C}_m(k) = u_1 \vec{P}(k) + u_2 \vec{P}(k+1) + u_3 \vec{P}(k-3),  
\end{equation}
where $k+1, k+4$ and $k-3$ are the tonic, major third and minor third respectively in the spiral representation. In the same fashion, the major key representations are defined from the major chords:
\begin{equation}
    \vec{T}_M(k) = \omega_1 \vec{C}_M(k) + \omega_2 \vec{C}_M(k+1) + \omega_3 \vec{C}_M(k-1),  
    \label{majkeys}
\end{equation}
where $k+1$ and $k-1$ are the dominant and the subdominant chords, the minor key representations are given by:
\begin{equation}
\begin{aligned}
\vec{T}_m(k) = {} & \nu_1 \vec{C}_m(k) \\ 
                  & + \nu_2 [ \alpha \vec{C}_M(k+1) + (1-\alpha) \vec{C}_m(k+1) ] \\
                  & + \nu_3 [ \beta \vec{C}_m(k-1) + (1-\beta) \vec{C}_M(k-1)],
\end{aligned}
\label{minkeys}
\end{equation}
this definition require two extra parameters to weight the different scale patterns that are presented in minor keys (natural, harmonic minor and melodic minor) given by the major and minor dominant and subdominant chords, the parameters $\alpha$ and $\beta$ are set equally: $$\beta = \alpha = 0.75.$$ 

Each weighting vector in the pitch, chord and keys definitions follow $w_1 \geq w_2 \geq w_3 > 0$ and $\sum_i w_i =1$, to consider some notes (or chords) more important than the others (e.g. the tonic in a chord would be more important than the fifth and both more important than the third). For convenience, all these vectors are given equal values: 
$$
\vec{\nu} = \vec{\omega} = \vec{u} = \vec{w} = \{0.536, 0.274, 0.19\} 
$$
 All parameters are discussed in details in the Appendix A (model calibration) of Chew's book about computational and mathematical modeling of tonality \cite{Chew2014}.
\begin{figure}[ht]
    \centering
    \includegraphics[width=0.6\columnwidth]{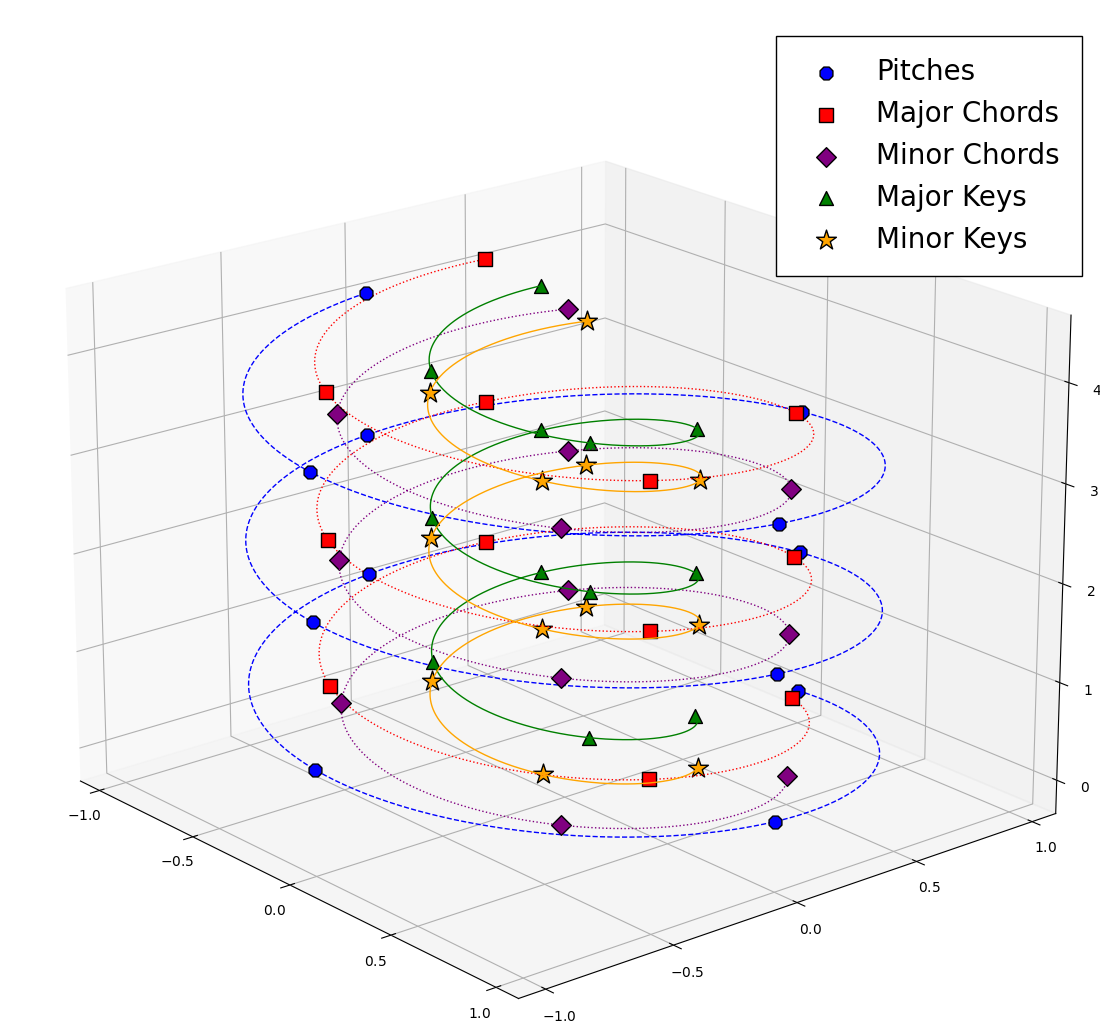}
    \caption{\textbf{Spiral representation.} Pitches, major and minor chords and keys in the geometric representation of the Spiral Array. }\label{spiralrep}
\end{figure}

\subsection*{Center of Effect}
The center of effect $\vec{C}_e$ is the location in $\mathbb{R}^3$ for the {\em center of mass} of a set of notes $P = \{ p_1, p_2, ..., p_N \}$ defined as a linear combination for the position of the pitches:
\begin{equation}
    \vec{C}_e = \sum_{i=1}^N \omega_i \vec{p_i} \label{coe1},
\end{equation}
where $\omega_i$ are the normalized ($\sum \omega_i = 1$) coefficients (weights) of each note, these coefficients can be defined in different ways since there are some notes that could be more important than others to define the key of the note sequence.

\subsubsection*{Duration weighting}
The first approach for the {\em importance} of a note is its duration, following the reasoning that a more lasting note will be more relevant for the harmony in the set of pitches. Given a set of duration $\{d_1,d_2, ...,d_N\}$, equation \ref{coe1} can be written as:
\begin{equation}
    \vec{C}_e = \frac{1}{\sum_i d_i}\sum_{i=1}^N d_i \vec{p_i} \label{coe2},
\end{equation}
where the term $\sum_i d_i$ is to guarantee that the coefficients are normalized. 

\subsubsection*{Pitch height weighting}
A different way of weighting the contribution of each note to the center of effect is to consider the height of the pitch: lower pitches are commonly used as the basis of the key (or chord) being normally the fundamental (tonic) note the most frequent in the low pitches. In the same fashion as the duration weighting, given a set of pitch height-weights $\{h_1, ..., h_n\}$, the center of effect can be re-written as: 
\begin{equation}
    \vec{C}_e = \frac{1}{\sum_i h_i}\sum_{i=1}^N h_i \vec{p_i} \label{coe3},
\end{equation}
where the weights $h_i$ can be assigned with a function of the height of the pitch, given a sequence of pitches $P = \{ p_1, p_2, ..., p_N \}$ each weight is computed by:
\begin{equation}
  h(p) =  1 + (\gamma -1) * (p - max(P)/(min(P)-max(P)),
\end{equation}
where the functions $max(P)$ and $min(P)$ correspond to the maximum and minimum value in the set $P$.

\subsubsection*{Beat weighting}
Other property that can be used for the coefficients of the center of effect is the number of beat on which the note is in the time signature of the measure. This feature consider the rhythmic feature since notes are not all the notes are equal in the measure, some notes are usually more emphasized than others depending on their location. Therefore depending on the time signature and the location of each note the equation \ref{coe1} can be written as:
\begin{equation}
    \vec{C}_e = \frac{1}{\sum_i b_i}\sum_{i=1}^N b_i \vec{p_i} \label{coe4},
\end{equation}

\subsubsection*{Multiple weighting}
Finally, the coefficients $\omega_i$ in eq. \ref{coe1} can be constructed as a combination of any of the weightings mentioned above, by combining all the tree weightings for the center of effect we obtain $\omega_i = d_i h_i b_i$.
\label{coe_def}

\section{CoE Key Finding Algorithm}
The CoE key finding algorithm uses the {\em Center of effect} as a reference to compute the {\em most likely key} for a sequence of notes. Given a sequence of pitches $P = \{\vec{p}_1, \vec{p}_2,...,\vec{p}_N\}$ its center of effect ($\vec{c}_e$) is computed, and the most likely key is given by:
\begin{equation}
    \argmin_{T \in \bm{T}} || \vec{c}_e - \vec{T} ||,
\end{equation}
which corresponds to the key $T$ for which the euclidean distance is minimum, where $\bm{T}$ is the set of major and minor keys: $\bm{T} = \{ \bm{T}_M(k) \forall k\} \cup \{\bm{T}_m(k) \forall k \}$ defined in equations \ref{majkeys} and \ref{minkeys}.  The associated likelihood of the ML key, as well as all alternative keys, is given by Eq.~\ref{eq:keylikelihood}.

\subsection*{Computing the center of effect with enharmonic equivalence}
Before computing the center of effects for the notes $\{ \vec{p}_1, ..., \vec{p}_n \}$ we make sure the notes follow the enharmonic equivalence assumption with the following steps:
\begin{itemize}
    \item 1) Map the notes in MIDI notation to a module 12 reference: $[0-127] \rightarrow [0-11]$, where the numbers $0-11$ correspond to the notes \{C,C\#,D,D\#/Eb,E,F,F\#/Gb,G,G\#/Ab,A,A\#/Bb,B\}.
    \item 2) Map the module 12 MIDI notation to the notes in the spiral array $\{ n_1, n_2, ..., n_n \} \rightarrow \{s_1, ...,s_n\}$ (ordered by fifths, with C$=0$).
    \item 3) Separate the notes in two groups $g_1 = \{s | s < 6 \}$ and $g_2 = \{ s | s >= 6\}$ and move one octave (12 units) all the elements of the group with less number of elements.  
    \item 4) For each element in $S = \{s_1, ...,s_n\}$ compute the difference $d(s_i) = |s_i - \bar{S}'|$, where $S'$ is all the elements in $S$ but the $i$th element. 
    \item 5) Move the element with the largest difference ($\argmax d(s_i)$) an octave towards the mean $\bar{S}'$ direction.
    \item 6) Repeat 4) and 5) until the average of $d(s_i)$ converges to a minimum.
    \item 7) Calculate the position for each note in the spiral array with equation \ref{pitchrep}.
\end{itemize}
These steps ensure that our final set of notes ($S$) follows the toroidal structure of the original Tonnetz model. 
\section{Local Keys to Modulation in Functional harmony}
\label{fun_har_sup}
We construct a sequence of roman numerals from a sequence of keys with the mapping $f: K\rightarrow H$ between the set of keys $K =$ \{C, G, D, A, E, B/Cb, Gb/F\#, Db/C\#, Ab, Eb, Bb, F, c, g, d, a, e, b, f\#, c\#, g\#/ab, eb/d\#, bb, f\} and roman numerals $H =$ \{I, I\#/IIb, II, II\#/IIIb, III, IV, IV\#/Vb, V, V\#/VIb, VI, VI\#/VIIb, VII, i, i\#/iib, ii, ii\#/iiib, iii, iv, iv\#/vb, v, v\#/vib, vi, vi\#/viib, vii\}, using the same musical notation as in \cite{Neuwirth2018}, where the key of reference (I) is the mode value for the distribution of keys in the piece or key sequence $\xi$: $K_I = mode(\xi)$. 

This mapping does not take into account the move to a different {\em center of harmony}, as a hand annotated functional harmony analysis would do. 
%This feature can be 
%interpreted as a downside but it also allows us to identify the exploration for the composer in the piece to different centers of harmony instead of taking each of them as the numeral I.
\label{app:subb3}
\\
\section{Validation of the CoE Algorithm}
We evaluate how accurate our implementation of the CoE (Center of Effect) algorithm performances by comparing the results we get from Beethoven's string quartets with annotated ones from professional musicians \cite{Neuwirth2018}. This annotated data is one of the most detailed set of musical analysis in the literature, it contains information about the key and chord in functional harmony terminology for each measure in all the Beethoven's string quartets. Our CoE implementation performs accurately in more than 68.43\% of the measures in the corpus provided by \cite{Neuwirth2018}.
\begin{table}[h]
\centering
\begin{tabular}{|l|l|l|l|l|l|}
\hline
\multicolumn{1}{|c|}{Measure} & \multicolumn{1}{c|}{Local\_Key} & \multicolumn{1}{c|}{Chord}                                         & \multicolumn{1}{c|}{Relative\_Numeral} & \multicolumn{1}{c|}{CoE\_KeyCall} & \multicolumn{1}{c|}{CoE\_Uncert} \\ \hline
1                             & I                               & .Eb.I                                                              & I                                      & I                                 & 0.27                                \\ \hline
2                             & I                               & V43                                                                & V                                      & I                                 & 1.54                 \\ \hline
3                             & I                               & I                                                                  & I                                      & I                                 & 0.28                                \\ \hline
4                             & I                               & V2                                                                 & V                                      & V                                 & 0.91               \\ \hline
5                             & I                               & I6                                                                 & I                                      & I                                 & 0.24                                \\ \hline
5                             & I                               & vi                                                                 & vi                                     & I                                 & 0.24                                \\ \hline
6                             & I                               & IV & IV                                     & IV                                & 1.15                 \\ \hline
7                             & I                               & IV                                                                 & IV                                     & ii                                & 0.38               \\ \hline
7                             & I                               & ii6                                                                & ii                                     & ii                                & 0.38                \\ \hline
7                             & I                               & V2                                                                 & V                                      & ii                                & 0.38                \\ \hline
8                             & I                               & I6                                                                 & I                                      & I                                 & 0.52                                \\ \hline
8                             & I                               & vi64                                                               & vi                                     & I                                 & 0.52                                \\ \hline
9                             & I                               & ii7                                                                & ii                                     & ii                                & 1.76                \\ \hline
9                             & I                               & V43                                                                & V                                      & ii                                & 1.76                 \\ \hline
10                            & I                               & I6(7)                                                              & I                                      & I                                 & 0.19                 \\ \hline
10                            & I                               & I6                                                                 & I                                      & I                                 & 0.19                 \\ \hline
10                            & I                               & @none                                                              &                                        & I                                 & 0.19                 \\ \hline
\end{tabular}\caption{Results of the CoE algorithm for key-calling compared to a hand-annotated corpus \cite{Neuwirth2018}. We show this comparison for the first ten measures of Beethoven's string quartet Op.~ 127. The hand-annotated key is denoted in the column entitled `` Relative\_Numeral". Some measures have more than one posisble key in the annotated corpus, whereas we show the most likely key per measure, according to the CoE algorithm.}
\end{table}
\label{app:supp4}
\newpage

\section{Diversity and Uncertainty distributions per historical period}
\begin{figure}[h!]
	\centering
    \includegraphics[width=0.9\textwidth]{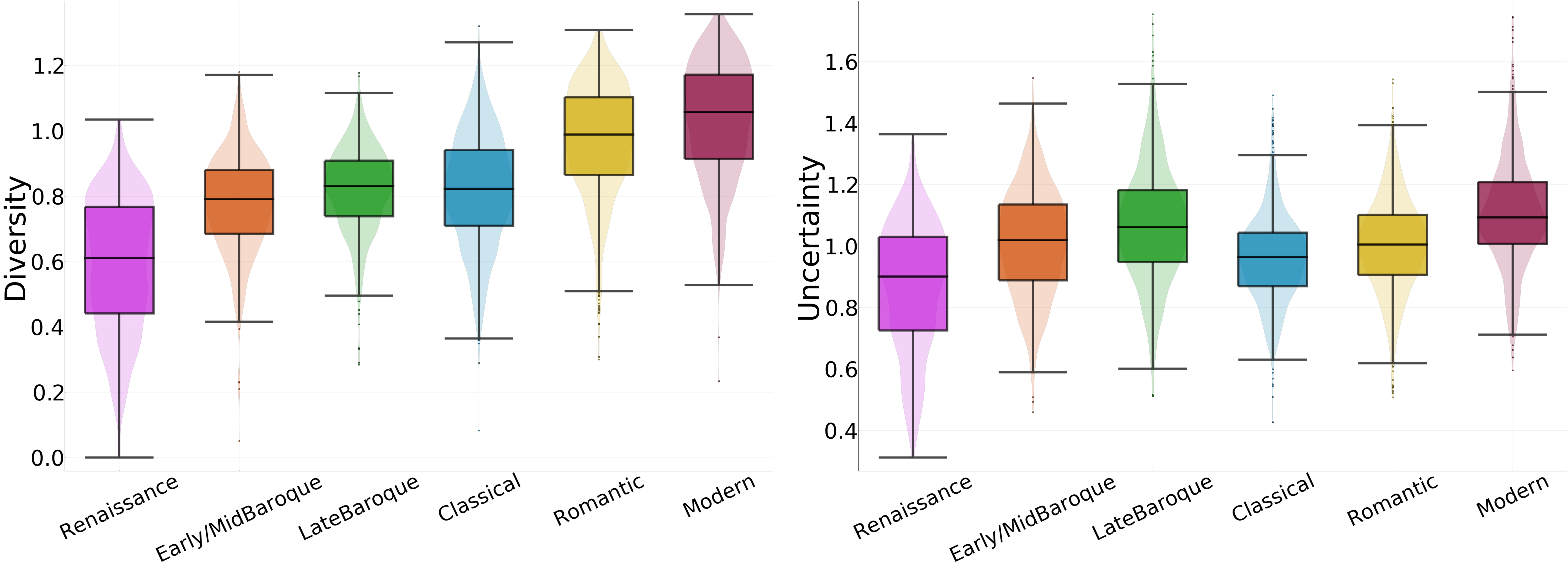}
    \caption{Diversity and uncertainty distributions on each historical period. Global trends from figures \ref{key_div} and \ref{key_unc} remain similar.}
	\label{bpdiv_unc}
\end{figure}
\newpage
\section{Selection of pieces per period with their novelty values}

\begin{longtable}{|c|c|c|c|c|}
\hline
Year & Name                        & Composer     & Novelty value & Sequence                                                                                                                                                                                                                                                                                                                                                                                                                                                                                                                                                                                                                                                                                                                                                  \\ \hline
1570 & O Sacrum Convivium          & F. Guerrero  & 0.2137        & \begin{tabular}[c]{@{}c@{}}I-IV-I-ii-I-IV-I-I-\\ V-IV-I-I-V-I-V-V-\\ I-IV-V-I-V-IV-I-V-\\ V-ii-I-IV-\#VI/bVII-I-\\ I-I-\#VI/bVII-IV-I-\\ I-I-I-V-V-I-I-vi-\\ I-vi-vi-vi-I-vi-I-\\ v-vi-ii-I-I-I-vi-\\ V-ii-ii-ii-vi-I-\\ v-IV-I-ii-I-I-\\ IV-v-I-V-I\end{tabular}                                                                                                                                                                                                                                                                                                                                                                                                                                                                                         \\ \hline
1696 & String Sonata II Mov 4      & D. Buxtehude & 0.1969        & \begin{tabular}[c]{@{}c@{}}I-I-IV-I-I-I-I-I-IV-I-I-vi-\\ V-I-I-I-V-I-V-I-V-vi-V-I-\\ vi-ii-I-vi-V-II-vi-V-V-V-\\ I-V-V-V-V-V-I-V-V-iii-\\ II-V-V-V-I-V-I-V-I-ii-I-IV-\\ ii-v-IV-ii-I-V-ii-I-I-I-\\ IV-I-I-I-I-I-IV-I-I-vi-V-\\ I-I-I-I-I-I-iii-I-vi-vi-ii-\\ vi-V-vi-V-V-V-V-I-I-\\ I-ii-V-V-I-I-IV-V-\\ vi-vi-I-vi-I-ii-V-V-I-\\ I-ii-I-V-I-I-IV-I-ii-V-I-\\ ii-iii-I-ii-I-ii-I-I-vi-III-\\ vi-iii-vi-iii-V-I-V-ii-vi-\\ VII-I-III-vi-V-I-I-IV-\\ i-v-ii-III-IV-iii-vi-IV-IV-\\ VII-III-vi-vi-iii-vi-vi-iii-\\ vi-I-iii-I-I-V-V-vi-ii-\\ vi-V-I-vi-vi-I-I-IV-IV-vi-\\ II-iii-I-vi-VII-iii-vi-IV-ii-\\ III-vi-ii-v-IV-IV-I-\\ ii-vi-vi-vi-IV-vi-v-vi-\\ IV-vi-vi-II-I-V-I-I-\\ vi-I-V-V-I-IV-V-IV-\\ I-IV-vi-v-v-II-V-I-\\ V-vi-I-I-IV-I-I-I\end{tabular} \\ \hline
1749 & Royal Fireworks Suite Mov 4 & G.F. Handel  & 0.1550        & \begin{tabular}[c]{@{}c@{}}I-I-I-I-V-V-V-V-I-\\ I-I-I-V-V-V-V-\\ V-V-I-I-I-I-V-\\ I-I-I-V-V-I-I-\\ I-I-V-I-I-I\end{tabular}                                                                                                                                                                                                                                                                                                                                                                                                                                                                                                                                                                                                                               \\ \hline

1798 & 7 Lander in D               & Beethoven    & 0.1670        & \begin{tabular}[c]{@{}c@{}}I-I-I-v-I-I-I-v-I-I-I-v-I-I-\\ I-v-I-ii-ii-ii-V-I-I-v-I-ii-ii-ii-V-I-I-v-\\ I-V-I-I-v-v-V-v-I-I-I-I-v-v-V-v-\\ I-I-ii-ii-ii-V-ii-ii-ii-V-I-I-v-v-V-\\ v-I-I-I-I-I-I-I-I-I-I-I-I-I-I-I-I-I-I-\\ V-V-V-v-I-I-I-I-V-V-V-v-I-I-I-I-I-\\ I-I-ii-I-i-I-v-I-I-I-ii-I-i-I-v-I-IV-\\ IV-V-I-IV-V-I-IV-IV-V-I-IV-\\ V-I-I-v-I-v-I-v-I-v-I-v-I-\\ v-I-v-I-v-I-i-IV-i-IV-v-I-v-I-\\ i-IV-i-IV-v-I-v-I-I-I-V-I-I-\\ I-V-I-I-I-V-I-I-I-V-I-I-\\ IV-I-v-I-IV-I-V-I-IV-I-\\ v-I-IV-I-V-I-I-I-I-I-\\ V-V-V-V-I-I-I-I-V-V-V-\\ V-I-IV-IV-IV-V-I-I-V-I-\\ IV-IV-IV-V-I-I-V-I-\\ I-I-i-I-I-I-I-I-I-I-I-\\ I-I-I-I-I-I-I-I-i-I-i-I-\\ I-I-I-I-I-I-I-I-I-I-i\end{tabular}                                                                            \\ \hline
1867 & Les Mois Op. 74 Mov 4       & Alkan        & 0.1888        & \begin{tabular}[c]{@{}c@{}}V-I-I-I-I-I-i-I-I-I-I-I-I-I-i-I-I-\\ I-I-I-I-I-I-I-I-I-I-I-I-I-I-I-I-I-\\ I-I-I-I-I-I-I-ii-V-ii-V-ii-\\ V-V-V-I-I-I-I-I-i-I-I-I-\\ I-I-I-I-I-I-I-V-V-I\end{tabular}                                                                                                                                                                                                                                                                                                                                                                                                                                                                                                                                                            \\ \hline
\caption{Full Key sequences for table \ref{low_inov}}
\end{longtable}

\begin{longtable}{|c|c|c|c|c|}
\hline
Year & Name                          & Composer    & Novelty value & Sequence                                                                                                                                                                                                                                                                                                                                                                                                                                                              \\ \hline
1570 & Madrigals Book 4 \#19         & C. Gesualdo & 1.72          & \begin{tabular}[c]{@{}c@{}}I-IV-I-vi-V-IV-I-IV-\\ ii-I-V-I-IV-V-V-I-\\ vii-V-ii-ii-vi-II-vi-\\ III-vi-I-V-vi-I-I-\\ vi-V-ii-V-II-II-vi-\\ III-vi-I-V-vi-I-I-\\ vi-V-ii-I-V-I-V-I-I\end{tabular}                                                                                                                                                                                                                                                                       \\ \hline
1690 & Mass for the parishes Mov 14  & F. Couperin & 1.8597        & \begin{tabular}[c]{@{}c@{}}V-vi-I-I-IV-I-I-IV-i-i-i-\\ v-IV-I-i-I-IV-\#II/bIII-i-\\ i-\#VI/bVII-iv-\#II/bIII-\#V/bVI-\\ \#II/bIII-\#VI/bVII-\#V/bVI-\\ \#VI/bVII-iv-i-\#VI/bVII-iv-\\ i-v-v-I-i-i-V-i-v-IV-v-II-\\ iv-iv-IV-I-v-iv-\#VI/bVII-\\ vi-\#II/bIII-\#VI/bVII-V-\\ ii-VI-V-V-\#i/bii-V-vi-ii-\\ \#V/bVI-i-VII-\#IV/bV\end{tabular}                                                                                                                           \\ \hline
1718 & WTC I Prelude 14 in F\# minor & J.S. Bach   & 2.1341        & \begin{tabular}[c]{@{}c@{}}i-i-\#VI/bVII-i-\#VI/bVII-\#II/bIII\\ -\#II/bIII-v-v-v-v-v-\#VI/bVII\\ -\#IV/bV-iv-i-i-i-i-i-\#II/bIII-iv-i-\\ I-\#II/bIII-ii-vii-\#VI/bVII-V-\\ II-i-i-I-i-i-v-i-iv-i-I-\\ VII-i-i-v-i-i-iv-i-\\ i-I-v-i-i-I-i-iv-i-\\ VII-IV-\#II/bIII-i-I-i-I\end{tabular}                                                                                                                                                                              \\ \hline
1817 & Sonata Op. 147 Mov 1          & F. Schubert & 1.6387        & \begin{tabular}[c]{@{}c@{}}I-I-I-V-I-I-I-I-IV-I-iv-\#II/bIII-\\ \#V/bVI-V-V-I-I-I-V-I-\\ I-V-iv-\#vi/bvii-V-I-i-v-ii-\\ \#V/bVI-\#II/bIII-\#V/bVI-III-\\ \#I/bII-\#i/bii-iv-VI-\#I/bII-\\ \#VI/bVII-i-\#V/bVI-\#II/bIII-\\ \#V/bVI-\#vi/bvii-\#V/bVI-\\ \#vi/bvii-\#V/bVI-\#ii/biii-\\ \#V/bVI-\#V/bVI-\#II/bIII-\\ I-I-I-V-I-I-I-I-IV-I-iv-\\ \#ii/biii-\#V/bVI-V-V-I-\\ IV-\#VI/bVII-\#II/bIII-\\ \#II/bIII-V-V-\#II/bIII-\\ \#II/bIII-I-IV-i-i-iv-I-I\end{tabular} \\ \hline
1861 & Esquisses Op. 63 No. 7        & Alkan       & 1.58          & \begin{tabular}[c]{@{}c@{}}I-I-I-I-vi-\#i/bii-v-\#i/bii-\\ III-III-III-\#i/bii-III-\#i/bii-\\ \#iv/bv-VI-\#i/bii-\#IV/bV-\\ vii-\#I/bII-\#V/bVI-i-\\ \#II/bIII-\#II/bIII-\\ i-\#V/bVI-V-i-i-\\ I-i-i-i-i-I-I\end{tabular}                                                          
\\ \hline
\caption{Full Key sequences for table \ref{hi_inov}}     
\end{longtable}

\label{full_seqs}

\section{Kullback-Leibler Divergence Rate}
To compute the Kullback-Leibler divergence between the two models ($\bm{P}_\xi$ for the piece and $\bm{P}_\Omega$ for the previous pieces) we can use the definition for the Entropy rate of a Markov Process in terms of its stochastic matrix $\bm{P}$ given by:
\begin{equation}
    H_R(\bm{P}) = - \sum_{ij} \mu_i P_{ij} log(P_{ij})
\end{equation}
where $\bm{\mu}$ is the asymptotic distribution of states in $\bm{P}$ and $P_{ij}$ are the probabilities for the transitions $i\rightarrow j$. From the relation of Entropy with Kullback-Leibler divergence, we know that:
\begin{equation}
    D_{KL}(P||Q) = H(P,Q) - H(P)
\end{equation}
where $H(P,Q)$ is the cross-entropy for distributions $P$ and $Q$ defined as $H(P,Q) = - \sum_{x \in \chi} P(x) log(Q(x))$, substituting the probability distributions with the stochastic matrices in terms of the Entropy Rate we obtain:
\begin{equation}
\begin{aligned}
    D_{KL_R}(\bm{P}||\bm{Q}) & = H_R(\bm{P},\bm{Q}) - H_R(\bm{P}) \\
                         & = -\sum_{ij}\mu_i P_{ij} log(Q_{ij}) - \sum_{ij}\mu_i P_{ij}log(P_{ij}) \\
                         & = \sum_{ij} \mu_i P_{ij} log\left(\frac{P_{ij}}{Q_{ij}}\right),
\end{aligned}
\end{equation}
substituting the values for the entries of the matrices $P$ and $Q$, we obtain:
\begin{equation}
    D_{KL_R}(\bm{P}_\xi||\bm{P}_{\Omega}) = \sum_{e_i \in V_\Xi} \sum_{e_j \in V_\Xi} \mu_{i} P_\xi(e_j|e_i)\cdot log\left(\frac{P_\xi(e_j|e_i)}{P_{\Omega}(e_j|e_i)}\right),
    \label{kld_rate}
\end{equation}
where the entries for $P_\xi(e_j|e_i)$ are defined in eq. \ref{prob_tran} and $P_\Omega(e_j|e_i)$ are computed as:
\begin{equation}
    P_{\Omega}(e_j | e_i) = \frac{f_{\Omega}(e_i,e_j) + \beta}{ \sum_{x \in V_\Xi} \left[f_{\Omega}(e_i, x) + \beta \right]},
    \label{p_icpp}
\end{equation}
the parameter $\beta$ is to avoid zero frequencies (when the transition $e_i \rightarrow e_j$ does not exist in $\Omega$) using additive smoothing with $ 0 < \beta < 1$. The quantity $D_{KL_R}(\bm{P}_\xi||\bm{P}_{\Omega})$ can be understood as the amount of information per time step necessary to reproduce the process distribution $\bm{P}$ from a model $\bm{Q}$.
\label{app:sup5}
\begin{figure}[h]
	\centering
    \includegraphics[width=0.8\textwidth]{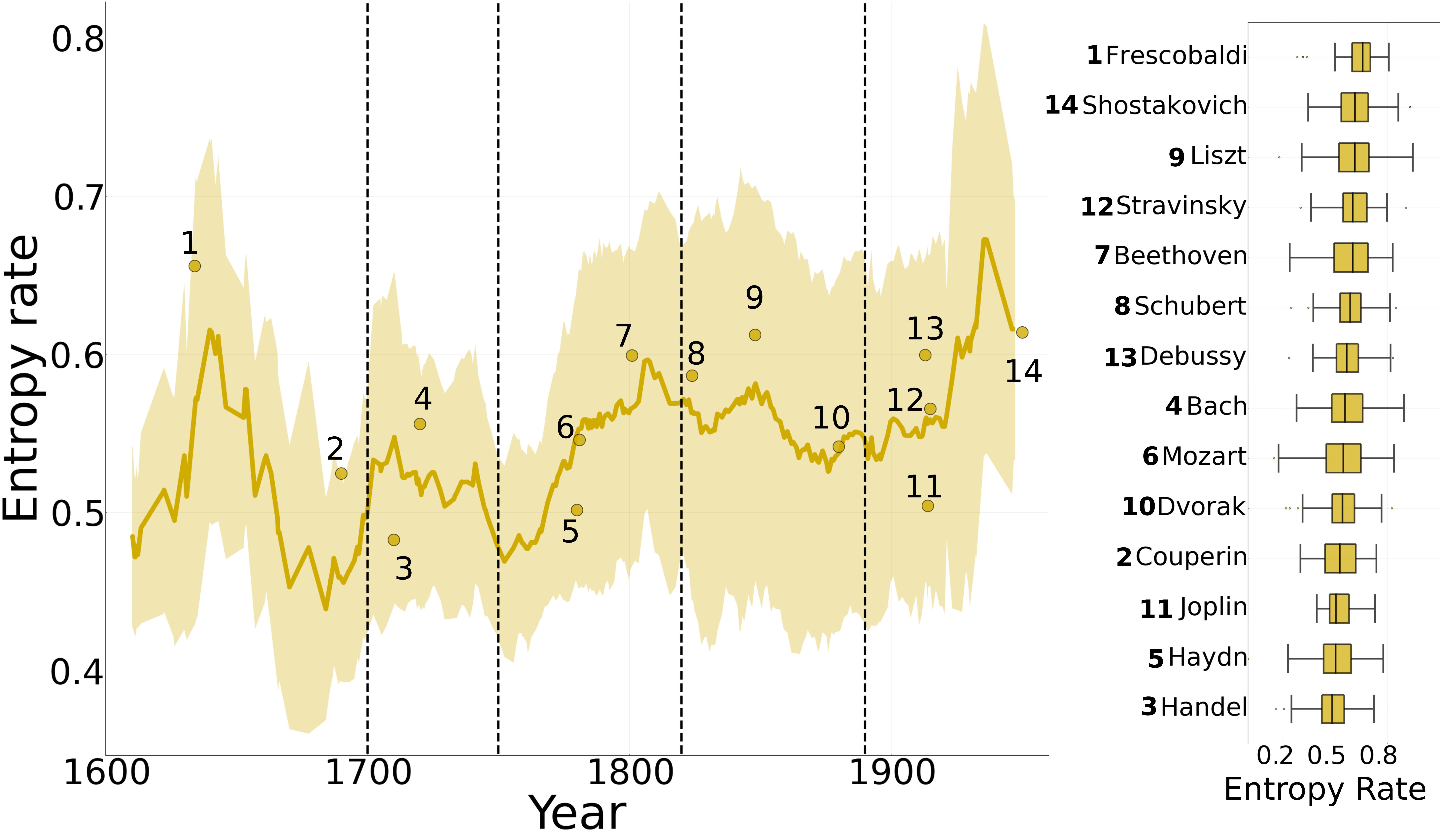}
    \caption{{\bf Entropy Rate.}}
	\label{ent_rate}
\end{figure}

\section{Dataset and data processing}
We processed both datasets with a Julia script (\textcolor{red}{ref}), extracting relevant information for the key calling algorithm such as: number of measure, starting time of the note, ending time and pitch (in MIDI notation).
 \label{app:supp1}
 
\begin{figure}[h]
	\centering
    \includegraphics[width=0.9\textwidth]{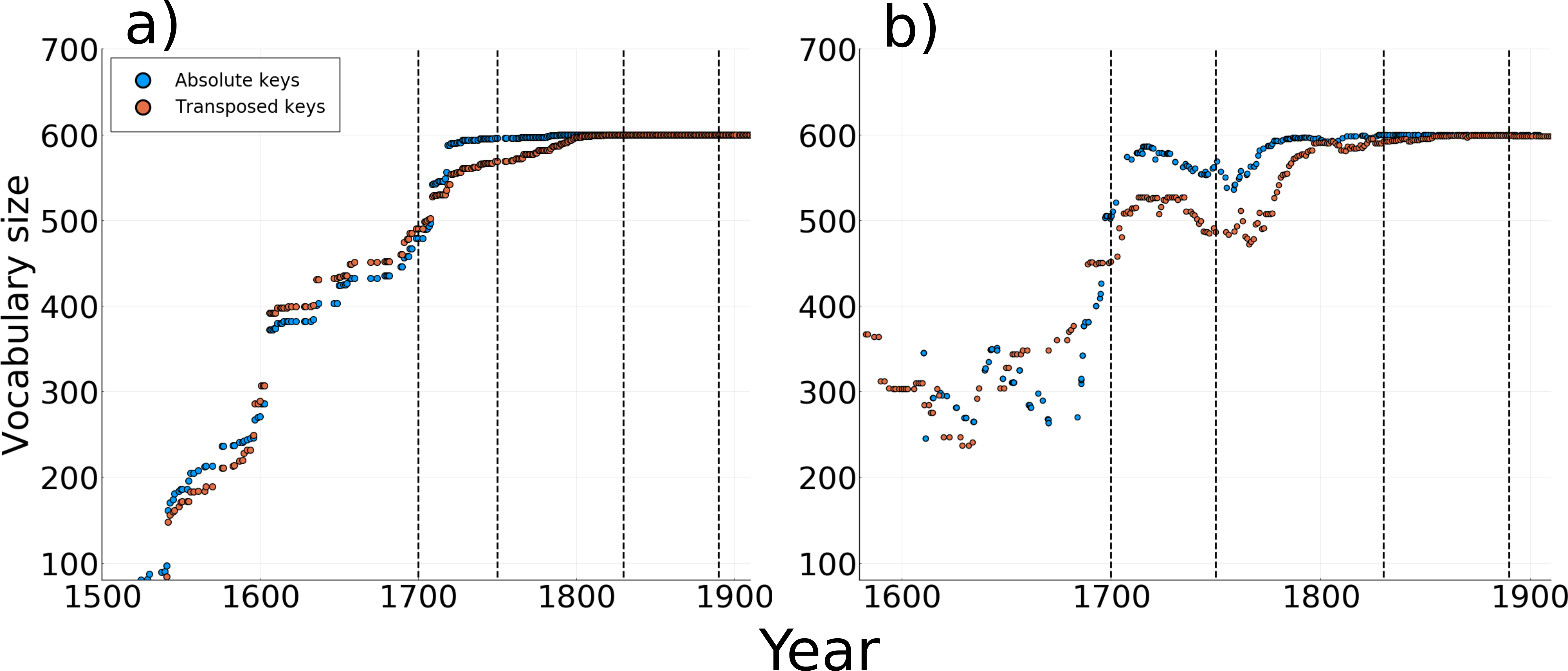}
    \caption{{\bf a) Cumulative vocabulary size. } {\bf b) Windowed vocabulary size.}  }
	\label{app:voc_size}
\end{figure}
\begin{figure}[h]
	\centering
    \includegraphics[width=0.9\textwidth]{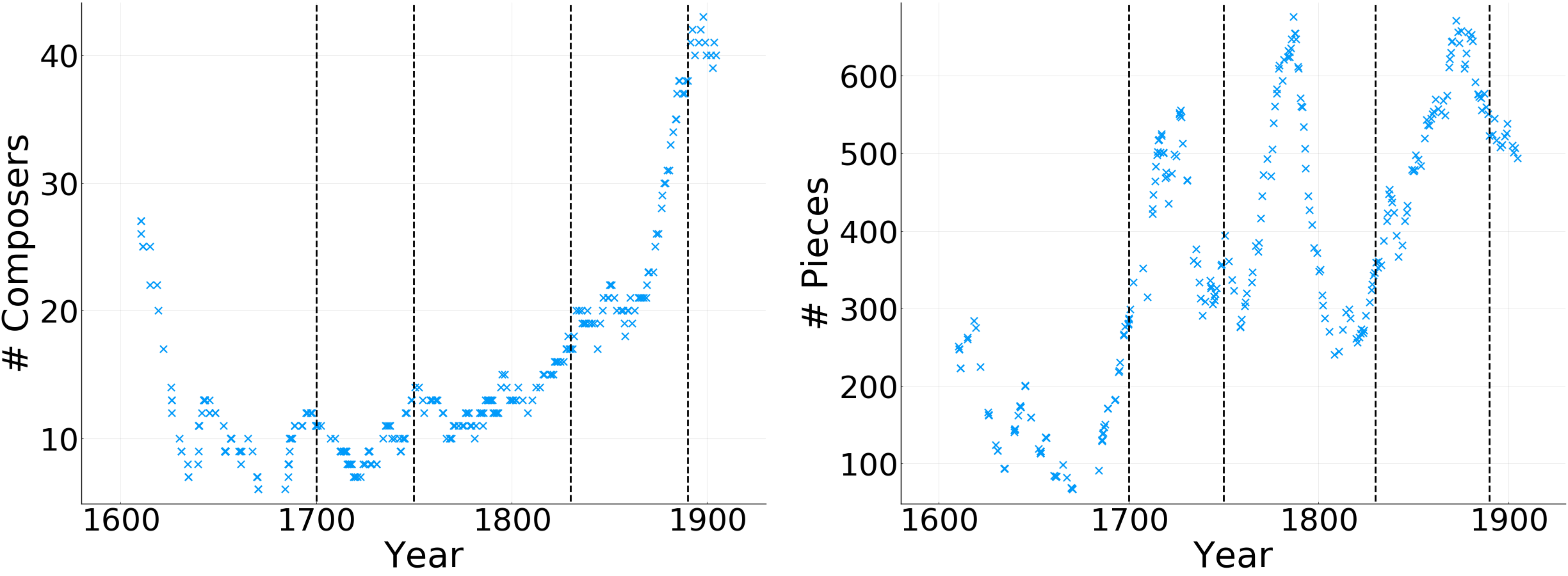}
    \caption{{\bf Left.} Number of different composers, over time. {\bf Right.} Number of pieces over the years.}
	\label{app:num_piec}
\end{figure}
\begin{figure}[h]
	\centering
    \includegraphics[width=0.6\textwidth]{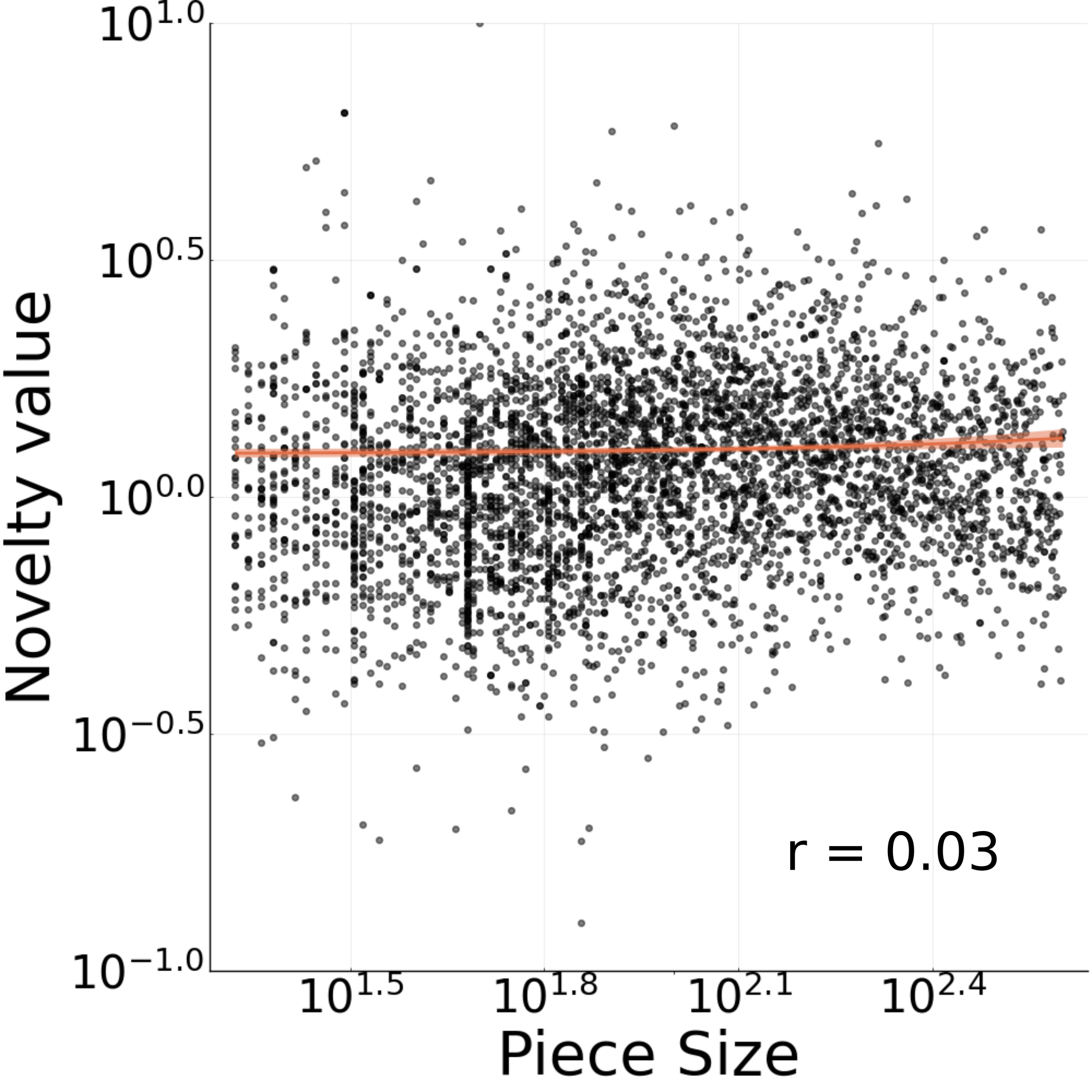}
    \caption{{\bf Novelty vs Size.}}
	\label{app:kld_size}
\end{figure}

 %%% Uncomment this line and comment out the ``thebibliography'' section below to use the external .bib file (using bibtex) .

\end{document}